\documentclass[10pt,twocolumn]{article}
\usepackage[top=2cm, bottom=2cm, left=1.8cm, right=1.8cm]{geometry}
\usepackage{times}  %
\usepackage[hyphens]{url}
\usepackage{graphicx}  %
\usepackage[keeplastbox]{flushend}
\usepackage{xcolor}
\usepackage{booktabs}
\usepackage{graphicx}
\usepackage{paralist}
\usepackage[small,bf]{caption}
\usepackage[tight]{subfigure}
\usepackage[hang,flushmargin]{footmisc}
\usepackage[compact]{titlesec}
\titlespacing*{\section}{0pt}{*4}{4pt}
\titlespacing{\subsection}{0pt}{*3}{3pt}
\usepackage{xspace}

\newcommand{\descr}[1]{\smallskip\noindent\textbf{#1}}

\newcommand{\td}{The\textunderscore Donald\xspace}

\begin{document}
\title{\textbf{Measuring and Characterizing Hate Speech on News Websites}}

\author{\bf Savvas Zannettou\textsuperscript{\rm 1}, Mai ElSherief\textsuperscript{\rm  2}, Elizabeth Belding\textsuperscript{\rm 3}, Shirin Nilizadeh\textsuperscript{\rm 4}, Gianluca Stringhini\textsuperscript{\rm 5}\\[0.5ex]
\normalsize \textsuperscript{\rm 1}Max Planck Institute for Informatics, \textsuperscript{\rm 2}Georgia Institute of Technology, \textsuperscript{\rm 3} University of California, Santa Barbara,\\ 
\normalsize \textsuperscript{\rm 4}University of Texas at Arlington, \textsuperscript{\rm 5}Boston University \\
\normalsize szannett@mpi-inf.mpg.de, melsherief@gatech.edu, ebelding@cs.ucsb.edu, shirin.nilizadeh@uta.edu, gian@bu.edu
}
\date{}

\maketitle

\begin{abstract}
The Web has become the main source for news acquisition. 
At the same time, news discussion has become more social: users can post comments on news articles or discuss news articles on other platforms like Reddit.
These features empower and enable discussions among the users; however, they also act as the medium for the dissemination of toxic discourse and hate speech.
The research community lacks a general understanding on what type of content attracts hateful discourse and the possible effects of social networks on the commenting activity on news articles.

In this work, we perform a large-scale quantitative analysis of 125M comments posted on 412K news articles over the course of 19 months.
We analyze the content of the collected articles and their comments using temporal analysis, user-based analysis, and linguistic analysis, to shed light on what elements attract hateful comments on news articles.
We also investigate commenting activity when an article is posted on either 4chan's Politically Incorrect board (/pol/) or six selected subreddits.
We find statistically significant increases in hateful commenting activity around real-world divisive events like the ``Unite the Right'' rally in Charlottesville and political events like the second and third 2016 US presidential debates.
Also, we find that articles that attract a substantial number of hateful comments have different linguistic characteristics when compared to articles that do not attract hateful comments.
Furthermore, we observe that the post of a news articles on either /pol/ or the six subreddits is correlated with an increase of (hateful) commenting activity on the news articles.
\end{abstract}

\section{Introduction}

As the Web becomes more social, so becomes the discourse around news events.
People share news articles on social media and discuss them with their friends~\cite{kwak2010twitter,zannettou2017web}. %
At the same time, news websites have become ``social,'' allowing users to post comments and discuss stories among themselves~\cite{diakopoulos2011towards,tsagkias2010news}.
While the ability to post comments empowers users to discuss news stories in a constructive fashion, discussion can also become toxic, leading to racist remarks and hate speech~\cite{erjavec2012you,harlow2015story,hughey2013racist}.
In particular, recent research showed that polarized Web communities such as 4chan's Politically Incorrect Board (/pol/) and Reddit's The\_Donald board often organize coordinated campaigns in which users are instructed to ``attack'' a target by using hate speech~\cite{flores2018mobilizing,hine2016kek,mariconti2018you}.
In some cases, these ``\emph{raids}'' can be directed towards news stories from sites that advocate policies that these users do not agree with.
Despite the problem that hate speech in news comments poses to news platforms and users, comment moderation remains an open problem~\cite{pavlopoulos2017deep}.

While hate speech and toxic discourse on social media has been the subject of study by a number of researchers~\cite{chatzakou2017mean,davidson2017automated,elsherief2018peer}, as a research community we still lack understanding on the characteristics and the dynamics of hateful comments on news articles.
In this paper, we perform a large-scale quantitative study of hateful news comments.
We analyze 125M comments from 412K news articles posted between July, 2016 and February, 2018.
To select the articles, we use all the news articles that are posted by popular news sites and for which links to them appear on 4chan's /pol/ and six selected subreddits from Reddit. 

\noindent \textbf{Research Questions.}
We aim to answer the following research questions:
1) Is hateful commenting activity correlated with real-world events?
2) Can we find important differences between the users that are posting on news sites according to their partisanship?
3) Can we find linguistic differences in articles that attract substantial numbers of hateful comments when compared to articles that do not? 
and
4) Do news articles attract more hate comments after they are posted on other Web communities like 4chan and Reddit?

To shed light on these research questions, we present a temporal and content analysis. 
We leverage changepoint analysis~\cite{killick2012optimal} to find significant changes in the time series of (hateful) commenting activity.%
We also use linguistic analysis that reveals the writing and linguistic peculiarities of news articles and whether articles that attract hate comments have differences to articles that do not attract hate.
Overall, this paper provides an unprecedented view on hateful commenting activity on news websites and on the characteristics of news articles that attract significant hate from users.

\noindent \textbf{Findings.} Among others, we make the following findings:
\begin{itemize}
    \item 
    We find a substantial increase in (hate) comments in close temporal proximity with important real-world events; e.g., we find statistically significant changes in hateful comments in news articles in close temporal proximity with the ``Unite the Right'' rally in Charlottesville during August, 2017, as well as the second and third US Presidential debates in 2016.
    \item We find differences between the users that are commenting on news articles according to the site's partisanship. Users that post on extreme-right sites tend to be more active overall by posting more comments and they tend to post more hateful content compared to users that are active on sites with other partisanships. Also, we find a higher percentage of hateful comments from users that choose to remain anonymous. 
    \item 
    Our linguistic analysis reveals that there is a correlation between articles using the highest number of Clout words (probably for influencing the readers) and attracting more hate comments. 
    We also find that the articles that had more than 10\% hateful comments, use more social references and include negative emotions, such as, \emph{anxiety} and \emph{anger} emotions, compared to those articles that receive no hate comment. 

    \item 
    We find a correlation between a link being posted on Reddit or /pol/, and receiving more (hateful) comments on that article. In particular, we find that the posting of news articles from domains with specific partisanships (i.e., Left, Center, Center-Right) to /pol/ or the six selected subreddits is correlated with an increase in hateful commenting activity in close temporal proximity with the posting of the news article on /pol/ or Reddit.
    We also discover that once a news article receives a substantial amount of hateful comments, it continues to receive a high fraction of such comments for a long period of time.
\end{itemize}

\section{Related Work}

\descr{Hate Speech Detection.}
A large body of work focuses on detecting hate speech.
HateSonar %
is a classifier~\cite{davidson2017automated} that uses Logistic Regression to classify text into: offensive language, or hate speech.
Recently, Google has released a state of the art hate speech detection tool, called Perspective API~\cite{jigsaw2018perspective}, that detects textual toxic content, including hate speech. This tool uses machine learning techniques and a manually curated dataset of texts, to identify the rudeness, disrespect, or toxicity of any comment. 
Most previous work~\cite{warner2012detecting,vigna2017hate,kwok2013locate,gao2017detecting,smedt2018automatic} proposes the use of supervised machine learning approaches, such as Support Vector Machines, Naive Bayes, and Logistic Regression, as well as Natural Language Processing techniques. 
Others~\cite{djuric2015hate,serra2017class,founta2018unified,gamback2017using} propose the use of neural network-based classifiers.%
Another work~\cite{gao2017recognizing} uses a semi-supervised approach to detect different forms of hate speech like implicit and explicit hate content.
Chandrasekharan et al.~\cite{chandrasekharan2017bag} propose Bag of Communities: an approach that uses data from 4chan, Voat, Reddit, and Metafilter, and aims to detect abusive content.
Finally, Saleem et al.~\cite{saleem2017awo} focus on multiple networks like Reddit and Voat, %
and propose the use of a community-driven detection approach. %

\descr{Hate Speech on the Web.}
Some recent work studies the prevalence and characteristics of hate speech on specific web communities, such as Gab~\cite{zannettou2018gab}, 4chan's Politically Incorrect board (/pol/)~\cite{hine2016kek}, Twitter and Whisper~\cite{silva2016analyzing}. 
Some works~\cite{mondal2017ams} study the effects of anonymity and forms of hate speech. 
Others~\cite{elsherief2018peer,Elsherief2018hate} perform an analysis on the personality of the targets and instigators of hate speech on Twitter. %
Another study by Zannettou et al.~\cite{finkelstein2018quantitative} shows the rise of racial slurs and in particular anti-semitism on 4chan and Gab.
Chandrasekharan et al.~\cite{chandrasekharan2017you} study the degree of hate speech on the platform after the bans of some prominent hateful subreddits like r/fatpeople and r/CoonTown, finding that these bans helped decrease the site's hate speech usage.
This is because a lot of accounts that were active on these subreddits stopped using the site and others that migrated to other subreddits did not post hateful content.
Olteanu et al.~\cite{olteanu2018effect} focus on understanding the effect that real-world extremist attacks, involving Arabs and Muslims, have on hateful speech on the Web.
Among other things, they observe an increase in the use of hate speech after such attacks and in particular increase in posts that advocate violence.
Jhaver et al.~\cite{jhaver2018online} study the effects of blocklists (i.e., blocking users) on online harassment, finding that users are not adequately protected online, while others feel that they are blocked unfairly.
Finally, a recent work by Zannettou et al.~\cite{zannettou2018origins} studies the dissemination of hateful memes across the Web. %

\descr{Hate Speech on News Comments.} 
Some studies analyze aspects of hate speech on comments posted on news articles.
Erjavec and Kovacic~\cite{erjavec2012you} undertake interviews with posters of hate speech on news sites to uncover their motives and strategies to share hateful content, finding that posters are driven by thrill and fun, while others are organized.
Hughey and Daniels~\cite{hughey2013racist} analyze the methodological pitfalls for studying racist comments posted on news articles. Specifically, they analyze various strategies employed by news platforms, such as extreme moderation policies, not storing comments or disabling comments, and their implications on the Web.
Harlow~\cite{harlow2015story} analyzes comments posted on US news sites to understand racist discourse.
They find that the comments included racial slurs despite the fact that the article did not;  Latinos were the most targeted ethnicity. %

\section{Methodology} \label{sec:method}
In this section, we describe our dataset collection process and our analysis methodology.
In a nutshell, we create a list of news sites, based on their popularity on 4chan's /pol/ and six selected subreddits, then we assess their partisanship, collect comments posted on their news articles whose links appear on /pol/ and the six subreddits, and finally, analyze their hate activity.

\begin{table}%
\centering
\resizebox{\columnwidth}{!}{%
\begin{tabular}{@{}llrrrr@{}}
\toprule
 & \textbf{Com. platform} & \textbf{\# of articles} & \textbf{\# articles} & \textbf{\# collected} & \textbf{\# collected} \\ 
\textbf{News site} & \textbf{(as of June 2018)} & \textbf{on /pol/} & \textbf{on 6 subreddits} & \textbf{articles} & \textbf{comments} \\ \midrule
dailymail.co.uk & Custom & 14,124 & 31,861 & 38,463 & 14,287,096 \\
theguardian.com & Custom & 10,430 & 49,318 & 42,137 & 11,090,592 \\
nytimes.com & Custom & 9,288 & 89,359 & 54,107 & 4,995,119 \\
washingtonpost.com & Custom & 9,213 & 136,120 & - & - \\
breitbart.com & Disqus & 7,698 & 39,793 & 41,918 & 46,684,682 \\
independent.co.uk & Custom & 6,232 & 28,971 & - & - \\
rt.com & Spot.IM & 5,980 & 13,913 & 17,075 & 2,707,512 \\
thehill.com & Disqus & 3,610 & 46,957 & 47,226 & 28,862,389 \\
almasdarnews.com & Oneall & 3,589 & 477 & - & - \\
express.co.uk & Spot.IM & 3,344 & 6,353 & 8,609 & 99,569 \\
huffingtonpost.com & Facebook & 3,009 & 34,999 & 27,092 & 1,089,113 \\
cbc.ca & Custom & 2,743 & 11,127 & - & - \\
dailycaller.com & Disqus & 2,727 & 18,516 & 19,457 & 5,326,962 \\
politico.com & Facebook & 2,684 & 26,247 & 19,916 & 626,386 \\
latimes.com & Custom & 2,091 & 15,902 & - & - \\
thesun.co.uk & Custom & 1,848 & 3,822 & - & - \\
washingtontimes.com & Spot.IM & 1,793 & 12,531 & 13,236 & 1,745,613 \\
mirror.co.uk & Custom & 1,734 & 5,001 & - & - \\
infowars.com & Disqus & 1,533 & 8,682 & 8,789 & 3,799,653 \\
newsweek.com & Facebook & 1,481 & 11,110 & 9,336 & 66,380 \\
sputniknews.com & Facebook+Custom & 1,380 & 3,808 & 4,343 & 29,368 \\
timesofisrael.com & Facebook & 1,301 & 4,367 & 4,588 & 110,466 \\
dailywire.com & Disqus & 1,173 & 6,892 & 7,343 & 603,208 \\
welt.de & Custom & 1,139 & 504 & - & - \\
jpost.com & Spot.IM & 1,080 & 4,037 & 4,707 & 294,250 \\
slate.com & Custom & 916 & 9,049 & - & - \\
salon.com & Spot.IM & 794 & 9,673 & 9,792 & 292,370 \\
huffpost.com & Facebook & 583 & 7,106 & 5,996 & 1,711,612 \\
townhall.com & Disqus & 548 & 7,015 & 7,235 & 693,372 \\
firstpost.com & Facebook & 76 & 23,310 & 20,759 & 555 \\ \midrule
\textbf{Total} &  & \textbf{104,141} & \textbf{666,820} & \textbf{412,124} & \textbf{125,116,267} \\ \bottomrule
\end{tabular}
}
\caption{Top news sources that support comments as of June, 2018, that appear on /pol/ and the six selected subreddits.} 
\label{tbl:news_sources_articles_counts}
\end{table}

\descr{Dataset.} Our dataset includes news articles and the comments posted on them between July 2016 and February 2018, on 4chan's Politically Incorrect board (/pol/) and six subreddits from Reddit, namely AskReddit, politics, conspiracy, \td, news, and worldnews.
We select these subreddits because they are among the most important subreddits when it comes to sharing news articles on Reddit%
~\cite{zannettou2017web}.
These subreddits attract both a general audience (i.e., news, politics,  worldnews, AskReddit subreddits), as well as users that are more into conspiracy theories and the alt-right (i.e., conspiracy, \td, and /pol/).
Due to this diversity in the Web communities where we collect news articles from, we expect that the collected articles will include a mixture of both mainstream, and possibly unbiased articles, as well as biased articles likely towards the alt-right community.

First, we extract all URLs that are posted on /pol/ and the six selected subreddits between July 2016 and February 2018. 
For obtaining the datasets for /pol/ we use the methodology presented by~\cite{hine2016kek}, while for Reddit we use publicly available data from Pushshift~\cite{baumgartner2020pushshift}.
Then, we select the top 100 domains according to their popularity in each online service. 
However, not every popular domain in these communities is actually a news site. 
For example, the most popular domain on /pol/ is YouTube~\cite{hine2016kek}.
Therefore, to identify domains that refer to \emph{news} sites, we used the Virus Total URL categorization API~\cite{virustotal}, which provides categories given a domain.
After obtaining the set of categories for each domain, we select the domains that have the ``news'' term in either of the returned categories, thereby obtaining a set of 64 news sites.
Then, during June 2018, we manually inspected these news sites to identify whether they allowed users to post comments, and if so what technology they used.
We found that 34 (53.1\%) sites do not support comments on their platform, six (9.3\%) sites use \emph{Disqus}~\cite{disqus}, five (7.8\%) sites use \emph{Spot.IM}~\cite{spotim}, seven (10.9\%) sites use \emph{Facebook}~\cite{facebook}, while twelve (18.7\%) sites use custom solutions.
The full list with all the sites is available at~\cite{full_list_sites}. %

Next, we aimed to implement tools to collect comments from the articles.
Initially, we looked at multiple domains that use the same commenting platforms; e.g., Disqus, Spot.IM, and Facebook.
For each of these, we built a crawler that uses the platform's API to get all the comments on articles posted on /pol/ or the six subreddits.
For news sites that use custom solutions as their commenting platforms, we had to implement a separate crawler for each domain, which is not efficient. %
Therefore, we focused on the domains for which we have the most articles; we implemented custom crawlers for \url{dailymail.co.uk}, \url{theguardian.com}, and \url{nytimes.com}. Note that we initially aimed to also implement a crawler for \url{washingtonpost.com} but we were unable due to implementation issues.
Table~\ref{tbl:news_sources_articles_counts} summarizes the number of the collected articles and comments for each news site that supports comments as of June 2018. 
Note that since we collect the data well after their publication date (collection period between June and November 2018), there is a small percentage of articles that are not available either because they were removed or because the URL was not available.
In total, we obtained 125M comments posted on 412K news articles.
Finally, for each article, we collected its content and associated article metadata using Newspaper3k~\cite{newspaper_library}.

\begin{table}[]
\centering
\caption{News sites in our dataset and their partisanship.}\label{tbl:partisanships}
\resizebox{\columnwidth}{!}{%
\begin{tabular}{@{}ll@{}}
\toprule
\textbf{Partisanship} & \textbf{News sites} \\ \midrule
Left & salon.com, huffpost.com, huffingtonpost.com, newsweek.com, firstpost.com \\ \midrule
Center-Left & nytimes.com, theguardian.com, thehill.com, timesofisrael.com \\ \midrule
Center & jpost.com, politico.com \\ \midrule
Center-Right & rt.com, washingtontimes.com, sputniknews.com \\ \midrule
Right & dailymail.co.uk, express.co.uk, dailycaller.com, dailywire.com, townhall.com \\ \midrule
Extreme-Right & breitbart.com, infowars.com \\ \bottomrule
\end{tabular}
}

\end{table}

\noindent \textbf{Identifying partisanship.}
To identify the partisanship of news sites, we use information about news media listed on the Media Bias/Fact Check (MBFC) website~\cite{mediabiasfactcheck}, which
contains annotations and analysis of the factual reporting and/or bias for news sites. MBFC has been used to annotate data in prior work for analyzing the factuality of reports and bias of news media~\cite{baly2018predicting}. Table~\ref{tbl:partisanships} shows the partisanship/bias of each news site in our dataset.

\noindent \textbf{Identifying hate comments.} To identify comments that are hateful, we explore the use of two popular hate speech classifiers: Hatesonar~\cite{davidson2017automated} and the Perspective API~\cite{jigsaw2018perspective}.
The former is a classifier that uses Logistic Regression to classify comments as hateful, offensive, or neither.
The classifier is trained on a corpus of 24K tweets annotated as either ``Hate Speech,'' ``Offensive Language,'' or ``Neither'' by workers on CrowdFlower.
Similarly, the Perspective API leverages crowdsourced annotations of text to train machine learning models that predict the degree of rudeness, disrespect, or unreasonableness of a comment.
In particular it offer two distinct models: the ``Toxicity'' and ``Severe Toxicity'' models.
The difference between the two models is that the latter is more robust to the use of swear words.
To assess the performance of these classifiers in our dataset, we extract a set of 100 random comments.
Then, three of the authors of this study independently marked each comment as hateful or not, and we treat the majority agreement of these annotations as groundtruth. 
Then, all comments in our random sample were evaluated both with HateSonar and the Perspective API. %
We find that HateSonar performs poorly on our random sample (precision 0.5 and recall 0.31), while the Severe Toxicity model of Perspective API performs substantially better (precision 0.71 and recall 0.52).
Interestingly, the Toxicity model of Perspective API performs better with respect to recall but is subpar in terms of precision (precision 0.53 and recall 0.84).
Based on these results, we elect to use the Severe Toxicity model available from Perspective API, mainly because we favor precision over recall and we aim to be more robust to the use of swear words (i.e., not everything that includes a swear word is hateful).

Note that hate speech detection is an open research problem and, to the best of our knowledge, there is no classifier that can detect all kinds and forms of hate speech.
This task is even difficult for humans as there are no clear definitions of what constitutes hate speech. 
For instance, in our random sample the three human annotators had a Fleiss Inter-Annotator agreement score of 0.39 that can be regarded as ``fair agreement''~\cite{wikipedia_fleiss}.
Due to this, in this work, we follow a best effort approach to study the prevalence and spread of hate speech using Perspective API that outperforms other readily available alternatives, such as the HateSonar classifier.

\section{Results}
In this section, we first provide a general characterization of the collected data with a focus on hateful content.
Next, we provide a user-based analysis to understand user activity on news article comments and then we investigate whether news articles with specific linguistic features attract more hateful content.
Finally, we examine whether there is any correlation between posting an article on 4chan's /pol/ and six subreddits and receiving hateful comments on those articles.

\subsection{General Characterization}
\begin{figure}[t!]
\center
\subfigure[]{\includegraphics[width=0.8\columnwidth]{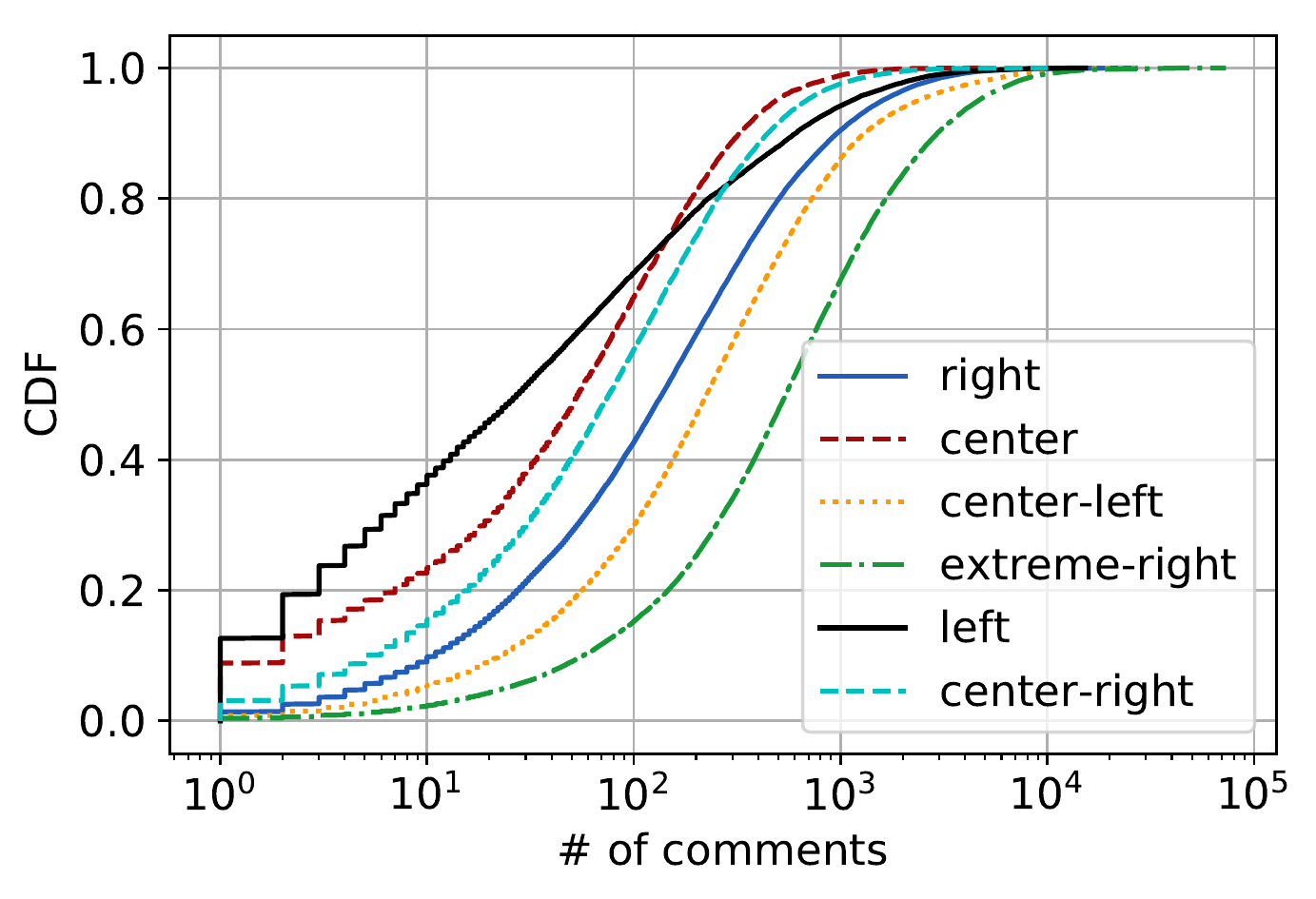}\label{subfig:cdf_all_comments_per_partisanship}}
\subfigure[]{\includegraphics[width=0.8\columnwidth]{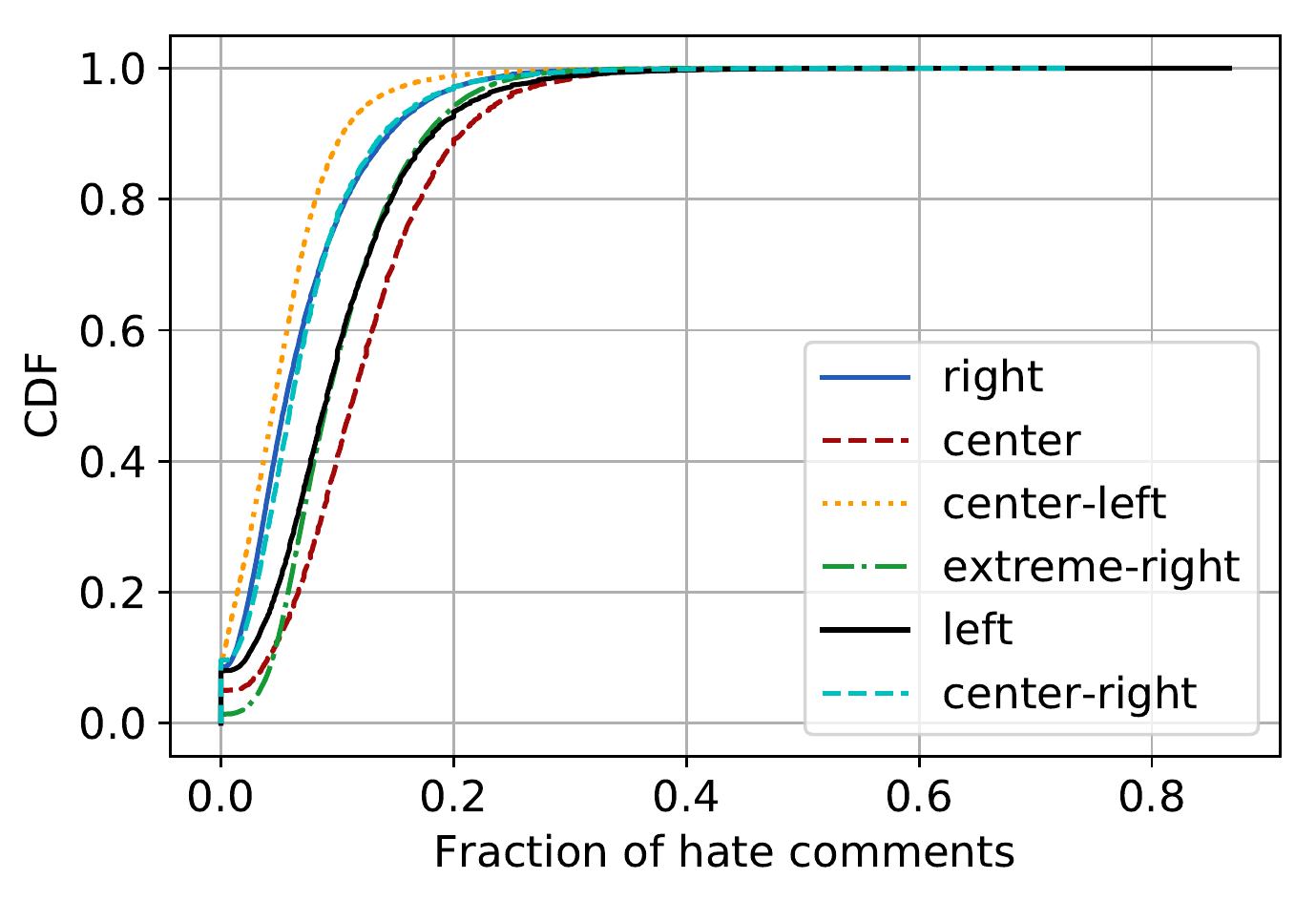}\label{subfig:cdf_hate_comments_per_partisanship}}
\caption{CDF of the number of (a) comments per article and (b) fraction of hate comments over all the comments per article.}
\label{fig:cdf_comments_per_platform}
\end{figure}

\descr{Prevalence of Hate Comments.} 
We present statistics of the comments that are posted for news articles and the prevalence of hate speech in these comments.
Fig.~\ref{fig:cdf_comments_per_platform} shows the cumulative distribution function (CDF) of the number of comments and the fraction of hate comments over all comments per news article, grouped by the partisanship of the news sites (see Table~\ref{tbl:partisanships}).
Note that for readability purposes we only show the distributions for articles that have at least one comment. 
When looking at the distribution of all the comments (Fig.~\ref{subfig:cdf_all_comments_per_partisanship}), we observe that extreme-right sites attract more comments, while left and center sites have a substantially lower commenting activity. 
To assess whether these results are affected by the different size of audiences for each news site, we use SimilarWeb~\cite{similar_web} to obtain the number of monthly views per news site (as of December 2018).
The full list of these views are publicly available via~\cite{list_views}.
Interestingly, we find that the most visited partisanship of news sites in our dataset is center-left (669M visits), followed by right (491M visits), center-right (286M visits), left (251M visits), extreme-right (77M visits), and last center (65M visits).
These findings indicate that the audience of left and extreme-right news sites are more active in posting comments despite the fact that center-left, right, and center-right news sites have a larger number of visits.

\begin{figure*}[t!]
\center
\includegraphics[width=0.7\textwidth]{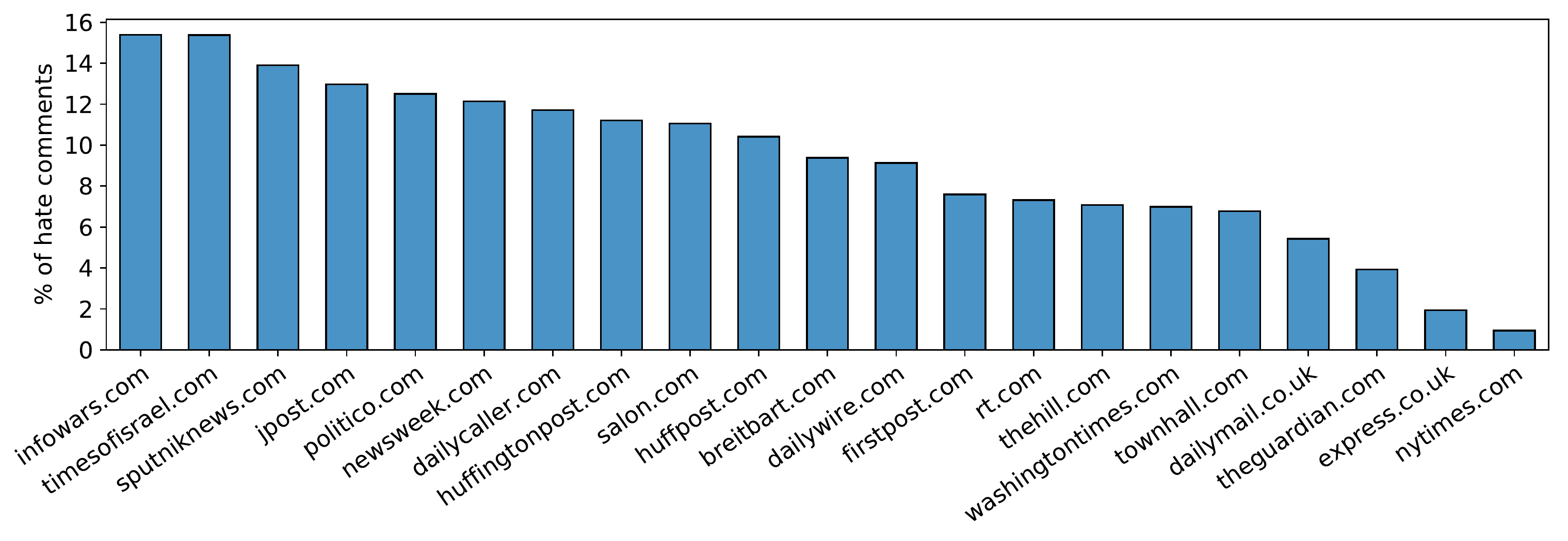}
\caption{Percentage of hate comments.} 
\label{subfig:bc_hate_comments_domain}
\end{figure*}

For hate comments (Fig.~\ref{subfig:cdf_hate_comments_per_partisanship}), we plot the fraction of hate comments over the overall number of comments per article. We find that center and left-leaning sites attract more hate speech, while center-left sites have the lowest rate of hate comments.
To assess whether the distributions shown in Fig.~\ref{fig:cdf_comments_per_platform} have statistically significant differences, we perform a two-sample Kolmogorov-Smirnov (KS) test for each pair of distributions; in all cases we find statistically significant differences with $p<0.01$.

Fig.~\ref{subfig:bc_hate_comments_domain} shows the percentage of hate comments over all the comments posted in news articles, grouped by news site.
We find that \url{infowars.com}, a popular alt-right conspiracy-oriented news site, and \url{timesofisrael.com} are the sites with the highest percentage of hate comments (15.3\%), followed by 
\url{sputniknews.com} (13.9\%), \url{jpost.com} (12.9\%), and \url{politico.com} (12.5\%).
When looking at the news sites with the least hateful commenting activity we find \url{nytimes.com} (0.9\%), followed by \url{express.co.uk} (1.9\%), and \url{theguardian.com} (3.9\%).
These results highlight the audience and comment moderation for each site: \emph{i.e.,} \url{infowars.com} is likely to attract users that post hate comments and the site might not apply strict moderation policies, while \url{nytimes.com} might  not attract hate comments or it might enforce strict moderation policies.
\begin{figure*}[t!]
\center
\subfigure[All comments]{\includegraphics[width=0.7\textwidth]{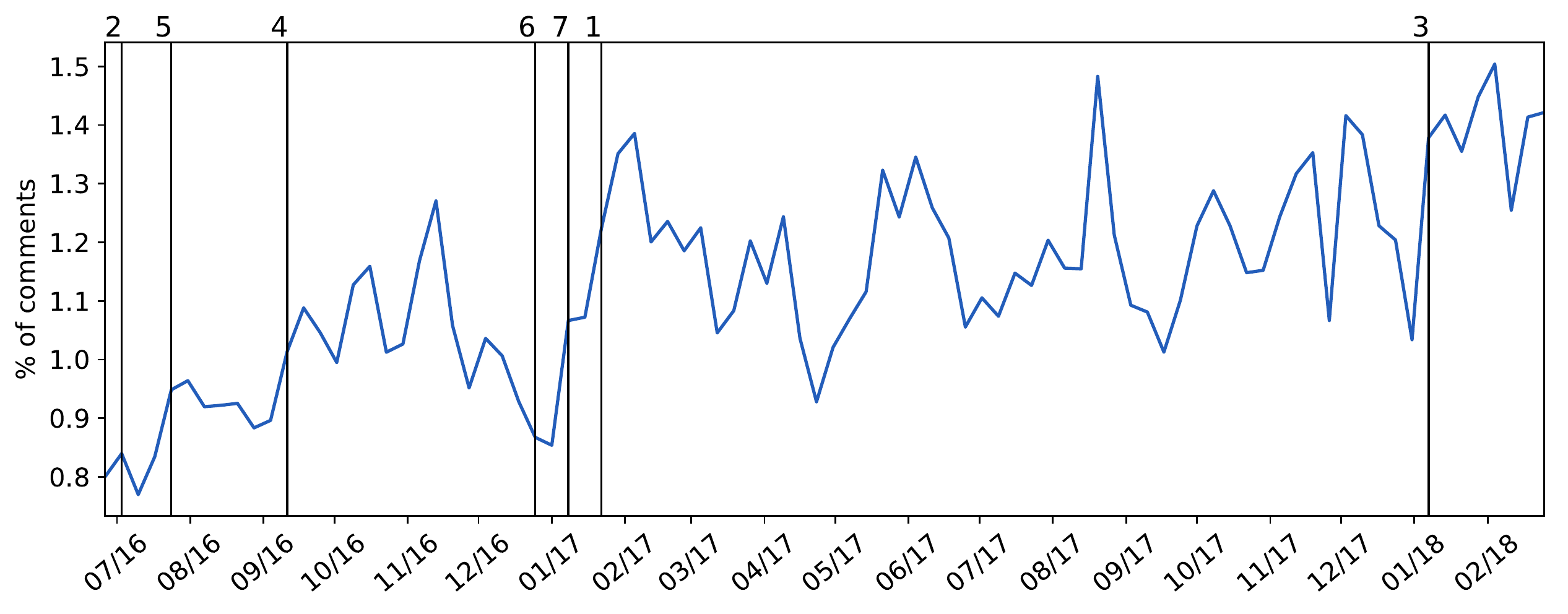}\label{subfig:temporal_all_comments}}
\subfigure[Hate comments]{\includegraphics[width=0.7\textwidth]{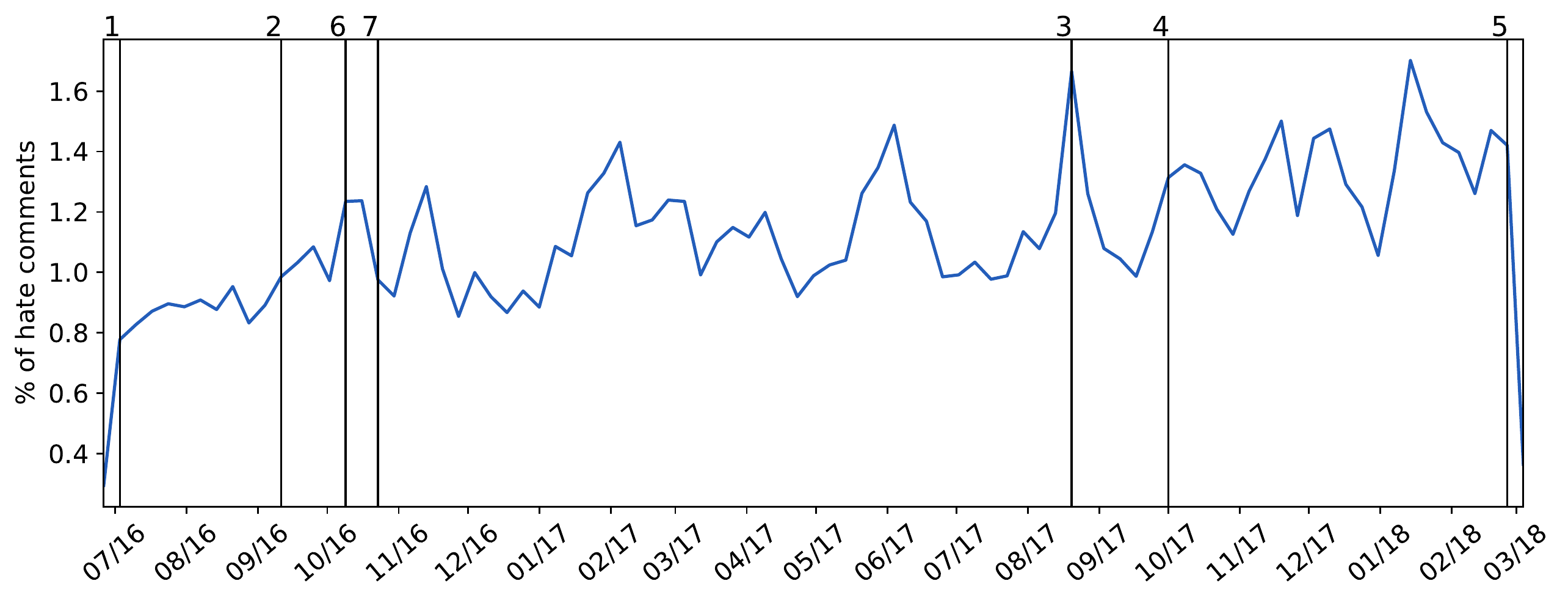}\label{subfig:temporal_hate_comments}}
\caption{Temporal overview of the collected comments. The figures are annotated with significant changes in the time series using changepoint analysis. See Tables~\ref{tbl:changepoints_all_comments} and~\ref{tbl:changepoints_hate_comments}, respectively, for real-world events that coincide with each changepoint.}
\label{fig:temporal_merged}
\end{figure*}

\begin{table*}[t!]
\centering
\resizebox{0.7\textwidth}{!}{%
\begin{tabular}{@{}ll@{}}
\toprule
\textbf{Changepoint} & \textbf{Events} \\ \midrule
1 - 2017/01/22 & 2017/01/20: Presidential Inauguration of Donald Trump~\cite{trump_inaguration}. \\ \midrule
2 - 2016/07/03 & 2016/07/02: Thousands of people protest in London against Brexit~\cite{brexit_protests}. \\ \midrule
3 - 2018/01/07 & 2018/01/02: Donald Trump responds to Kim Jong-Un stating that his nuclear missile launch button\\  
               & is larger and more powerful~\cite{trump_button}. \\ \midrule
& 2016/09/09: US congress passes a law to allow families of 9/11 victims to sue Saudi Arabia~\cite{sue_saudi_arabia}.\\   
4 - 2016/09/11  & 2016/09/11: Hillary Clinton is treated for pneumonia after leaving a ceremony  honoring \\  
  & the anniversary of 9/11 attacks~\cite{clinton_sick}.\\\midrule
5 - 2016/07/24 & 2016/07/19: Donald Trump is nominated as the Republican's candidate for the 2016 US election~\cite{trump_nomination}. \\\midrule
6 - 2016/12/25 & 2016/12/22: Donald Trump names Kellyanne Conway as Counselor to the President\\  
 & and Sean Spicer as White House Press Secretary~\cite{counsellor_trump,wh_secretary}.\\\midrule
7 - 2017/01/08 & 2017/01/06: A US intelligence document reports that Vladimir Putin ordered a campaign to influence\\  
 & the 2016 US election~\cite{putin_us_elections}. \\ \bottomrule
\end{tabular}%
}
\caption{Statistically significant changepoints and coinciding real-world events in the time series of all the comments.}\label{tbl:changepoints_all_comments}
\end{table*}

\begin{table*}[t!]
\centering
\resizebox{0.7\textwidth}{!}{%
\begin{tabular}{@{}ll@{}}
\toprule
\textbf{Changepoint} & \textbf{Events} \\ \midrule
1 - 2016/07/03 & 2016/07/02: Thousands of people protest in London against Brexit~\cite{brexit_protests}. \\\midrule
& 2016/09/09: US congress passes a law to allow families of 9/11
victims to sue Saudi Arabia~\cite{sue_saudi_arabia}.\\   
2 - 2016/09/11 & 2016/09/11: Hillary Clinton is treated for pneumonia
after leaving a ceremony honoring\\  
 & the anniversary of 9/11 attacks~\cite{clinton_sick}. \\ \midrule
3 - 2017/08/13 & 2017/08/11: Unite the Right rally begins in Charlottesville, Virginia~\cite{charlotesville}. \\ \midrule
4 - 2017/10/01 & 2017/10/02: Shooting in Las Vegas leads to the death of 59 people~\cite{las_vegas_shooting}. \\ \midrule
5 - 2018/02/18 & 2018/02/14: Shooting at Stoneman Douglas High School with 17 people dead~\cite{florida_shooting}. \\ \midrule
6 - 2016/10/09 & 2016/10/09: Second presidential debate of the 2016 US election take place~\cite{second_debate} \\ \midrule
7 - 2016/10/23 & 2016/10/19: Third presidential debate of the 2016 US election take place at Las Vegas~\cite{third_debate}. \\ \bottomrule
\end{tabular}%
}
\caption{Statistically significant changepoints and coinciding real-world events in the time series of hateful comments.} 
\label{tbl:changepoints_hate_comments}
\end{table*}

\descr{Temporal Analysis.} 
Here, we examine the temporal aspect of the collected comments to understand how (hateful) commenting activity changes over time.
This is a particularly interesting and important analysis since it will allow us to understand whether hateful commenting activity is correlated with real-world events and whether hateful commenting activity is increasing or decreasing over time.
Fig.~\ref{fig:temporal_merged} shows the weekly percentage of comments and hateful comments for the whole dataset. 
We focus on the time period after July, 2016, as the vast majority of the collected comments are within the depicted time period. 
We find that the overall commenting activity started increasing during the months leading to the 2016 US election (between September and November 2016), decreased after the election, while again started increasing after Trump's Inauguration (January 2017). 
Furthermore, we note that the biggest peak in commenting activity coincides with the ``Unite the Right'' rally in Charlottesville~\cite{charlotesville}, during August 2017, which lead to the death of one woman~\cite{charlotesville_death}.
When looking at the hate comments (Fig.~\ref{subfig:temporal_hate_comments}), we find a somewhat similar activity with all the comments (Fig.~\ref{subfig:temporal_all_comments}).
Some peaks in hateful commenting activity coincide with the 2016 US election period, with Trump's Inauguration in January 2017, with the Charlottesville rally in August 2017.
Since our dataset is based on articles posted on 4chan's /pol/ and the six subreddits, these findings indicate that their users are particularly interested in discussing these political events and that they likely comment on them both on their platform as well as in the comments section of each article. 

We further investigate whether the peaks in overall and hate commenting activity are statistically significant with respect to the time series of the comments. 
We run changepoint analysis that provides points in time where statistically significant changes occur on a time series. 
Specifically, we run the Pruned Exact Linear Time (PELT) algorithm~\cite{killick2012optimal} on the weekly time series of both all comments and hate comments.
This algorithm maximizes the log-likelihood of the means and variances of the time series with a penalty function that enables us to rank the changepoint according to their statistical significance.
Fig.~\ref{fig:temporal_merged} is annotated with the obtained changepoints for both all comments and hate comments, while Tables~\ref{tbl:changepoints_all_comments} and~\ref{tbl:changepoints_hate_comments} report each changepoint and real-world events that coincide with each changepoint.  
For the overall commenting activity we find statistically significant changepoints that coincide with the Presidential Inauguration of Donald Trump (changepoint 1 in Table~\ref{tbl:changepoints_all_comments}), Brexit protests (changepoint 2 in Table~\ref{tbl:changepoints_all_comments}), and developments on the USA-North Korea relations (changepoint 3 in Table~\ref{tbl:changepoints_all_comments}).
For hateful commenting activity we find statistically significant changepoints that coincide with Brexit developments (changepoint 1 in Table~\ref{tbl:changepoints_hate_comments}), the Las Vegas shooting during October 2017 (changepoint 4 in Table~\ref{tbl:changepoints_hate_comments}), developments in US politics (changepoint 2 in Table~\ref{tbl:changepoints_hate_comments}), and the presidential debates during the 2016 US election (changepoints 6 and 7 in Table~\ref{tbl:changepoints_hate_comments}).  Finally, we find a changepoint coinciding with the Charlottesville protest (changepoint 3 in Table~\ref{tbl:changepoints_hate_comments}).

\subsection{User Analysis}
In this section, we analyze the users that comment on news articles.
We are particularly interested in understanding how these users interact in the comments of news articles, how persistent users are in disseminating hateful comments, and whether users that post on news sites with specific partisanship are more hateful.
Furthermore, since some commenting platforms (e.g., Disqus) allow users to post comments anonymously, we investigate the effect of anonymity with respect to the dissemination of hateful comments on news articles.
Note that due to ethical reasons, we do not make any attempt to link users across the multiple commenting platforms we study, while at the same time we make no attempt to de-anonymize users.

\descr{Effect of anonymity.} We investigate the prevalence of  posting  comments anonymously.
We find that in our dataset 6.5M (5.2\%) comments are posted by anonymous users, while the rest of the comments are posted by users that have accounts with the various commenting platforms we study.
Next, we look into the prevalence of hateful comments in each of these subsets: we find that in the anonymous subset there are relatively more hateful comments (10.7\% of them), while for the subset where users had accounts we find a lower percentage of hateful comments (7.6\%), which is inline with previous work focusing on hate speech on anonymous and non-anonymous platforms~\cite{mondal2017measurement}.
We also assess the statistical significance of these results with a Chi-square test on the number of hateful and non-hateful comments for anonymous and non-anonymous users, finding statistically significant differences ($p<0.01$). 
Overall, these findings indicate that most users do not mind creating an account on these commenting platforms and that users that choose to remain anonymous are more likely to share hateful comments.

\begin{figure}[t!]
\center
\subfigure[All users]{\includegraphics[width=0.8\columnwidth]{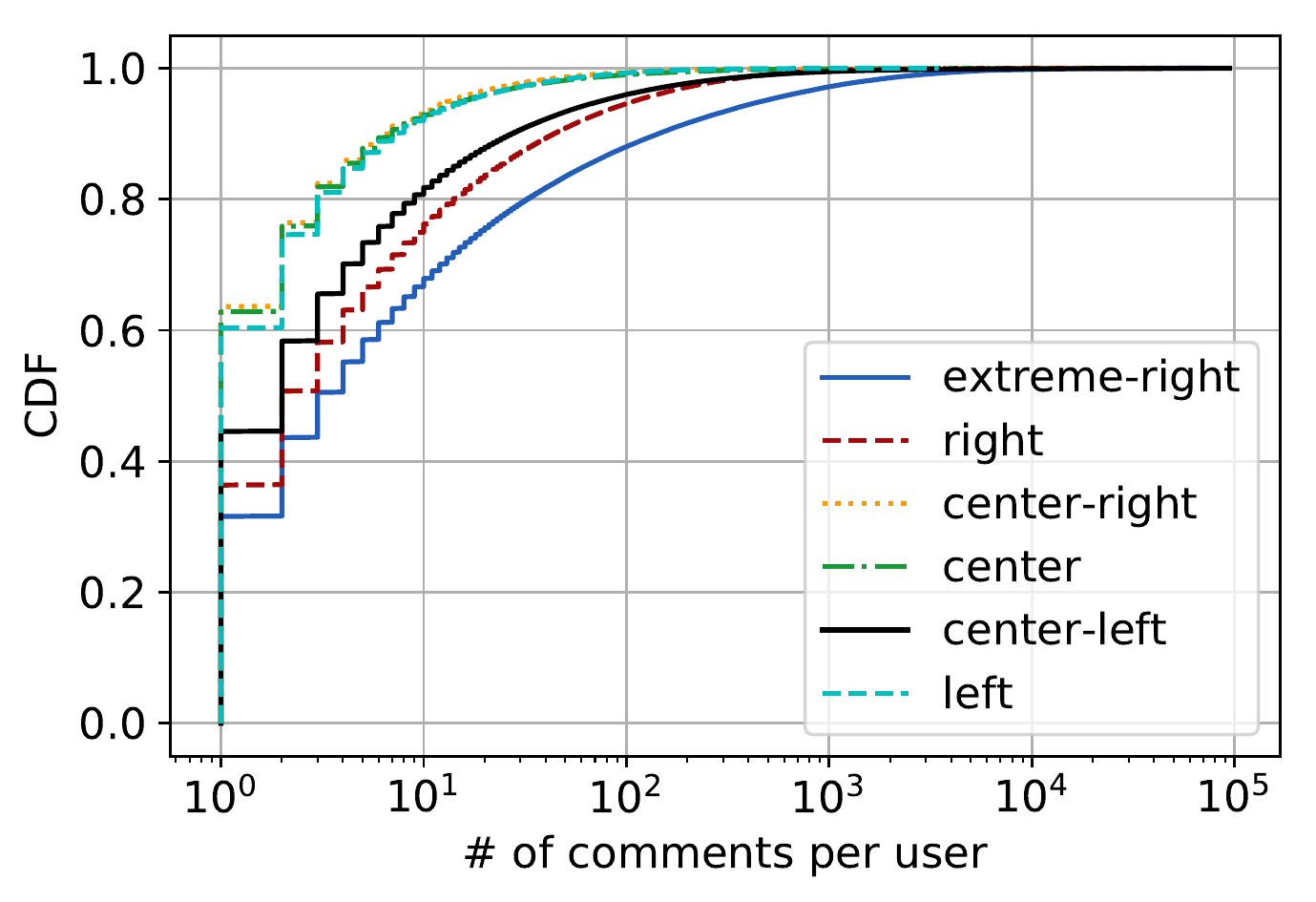}\label{subfig:cdf_comments_per_user}}
\subfigure[Users~with~at~least~ten~comments]{\includegraphics[width=0.8\columnwidth]{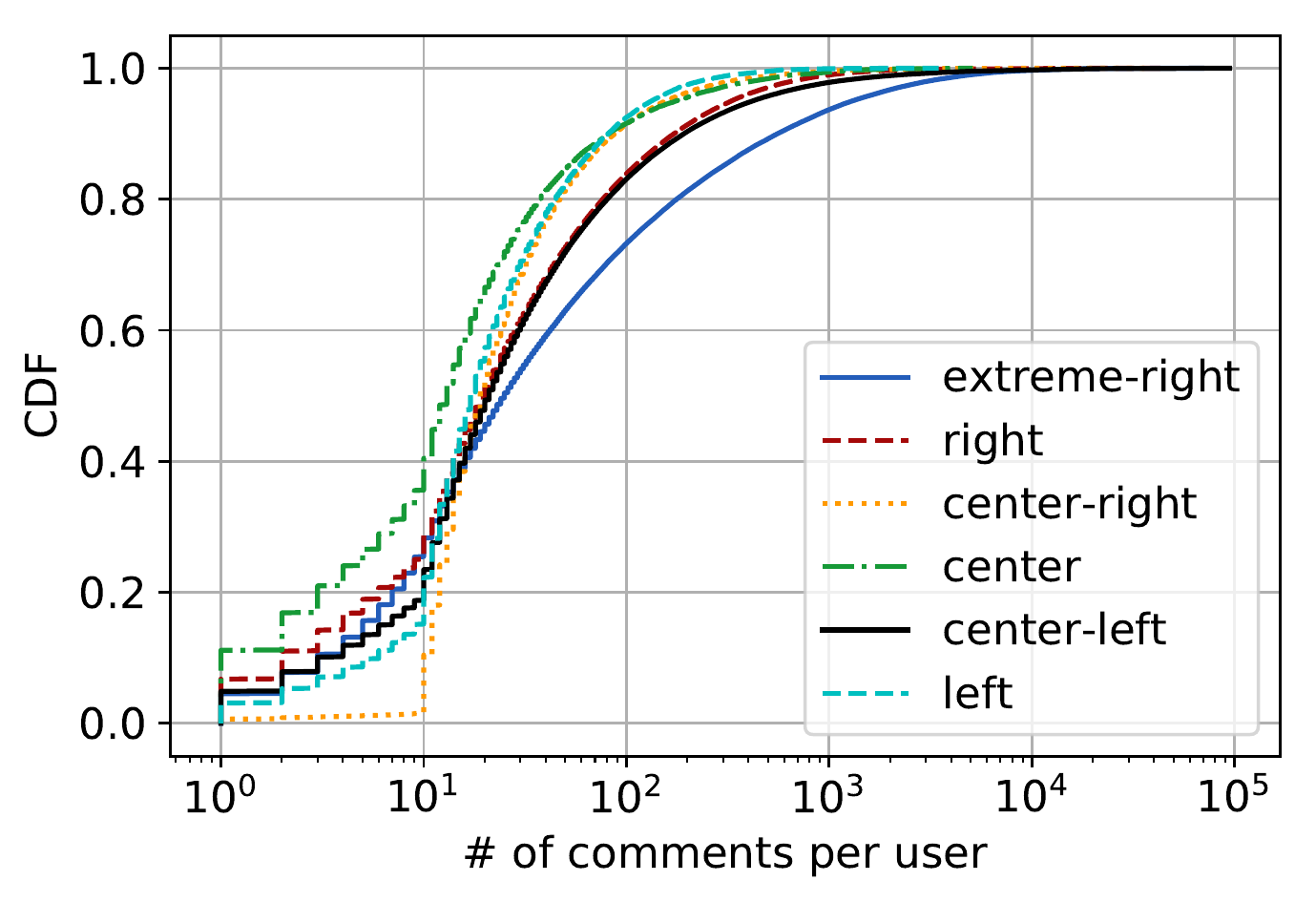}\label{subfig:cdf_comments_per_user_more_10}} \caption{CDF of the number of comments per user.}
\label{fig:cdf_comments_per_user}
\end{figure}

\descr{Overall User Activity.} Since we want to analyze the dataset in the granularity of specific users, we therefore next focus on the subset of the dataset where users posted comments by creating accounts on the commenting platforms.
Overall, we find 3.1M accounts across all the commenting platforms. 
To get a better understanding of how users interact with news comments, we plot the CDF of the number of comments per user in our dataset in Fig.~\ref{fig:cdf_comments_per_user}. 
Since a substantial percentage of users had only posted one comment,  we show the results for users that posted at least ten comments through all the articles in Fig.~\ref{subfig:cdf_comments_per_user_more_10}.
Specifically, we find that from the users that are active on extreme-right news articles comments, 31\% of them posted only once across all news articles, while the same percentage increases for other partisanships: 36\% for right, 44\% for center-left, 60\% for left, and 63\% for center and center-right.
Furthermore, we note that users that post on extreme-right news articles comments are more active (mean number of comments 134.32 ), followed by users on center-left (mean number of comments 38.6) and right (mean 29.9). 

\begin{figure}[t!]
\center
\subfigure[All users]{\includegraphics[width=0.8\columnwidth]{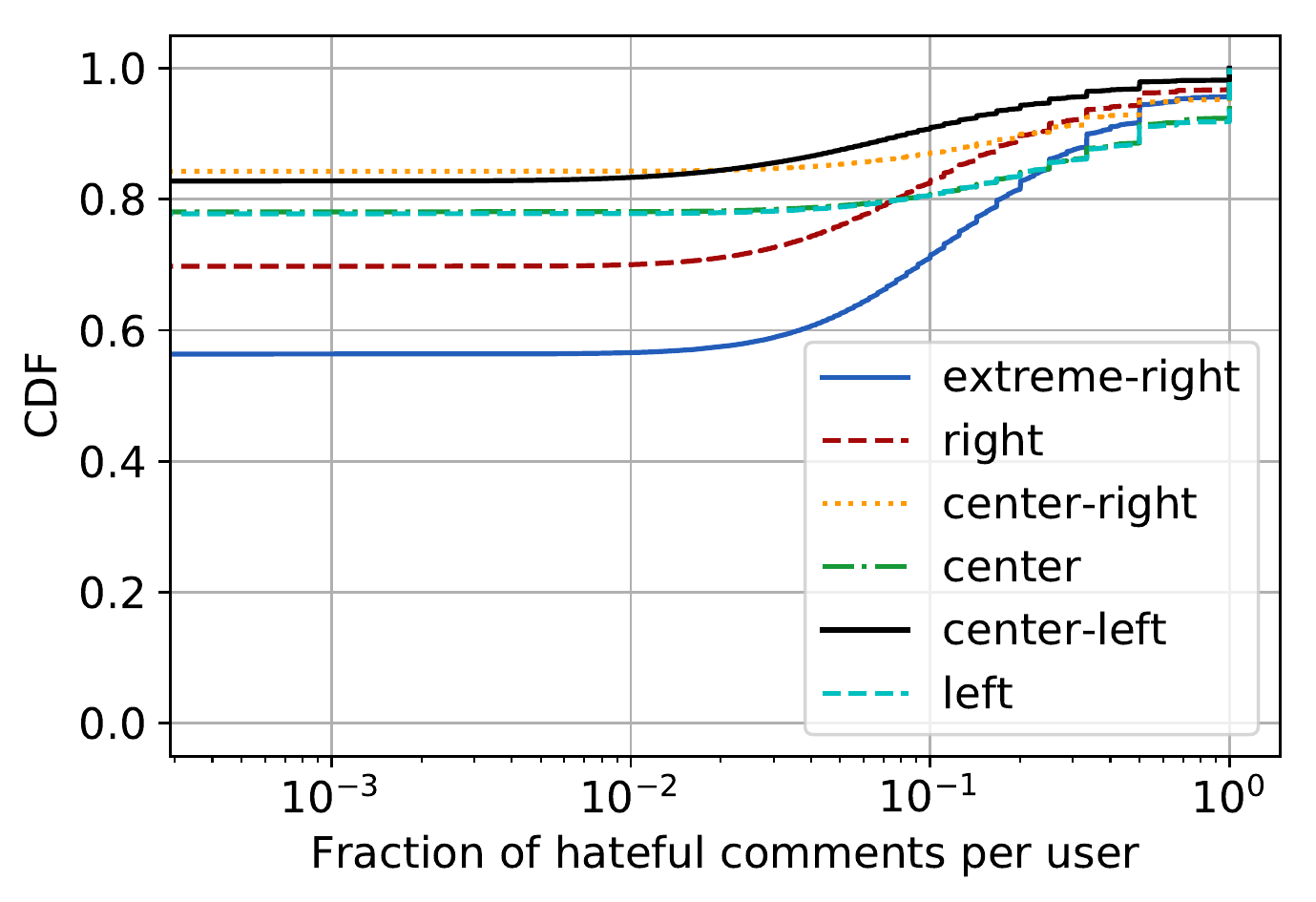}\label{subfig:cdf_fraction_hateful_comments_per_user}}
\subfigure[Users~with~at~least~ten~comments]{\includegraphics[width=0.8\columnwidth]{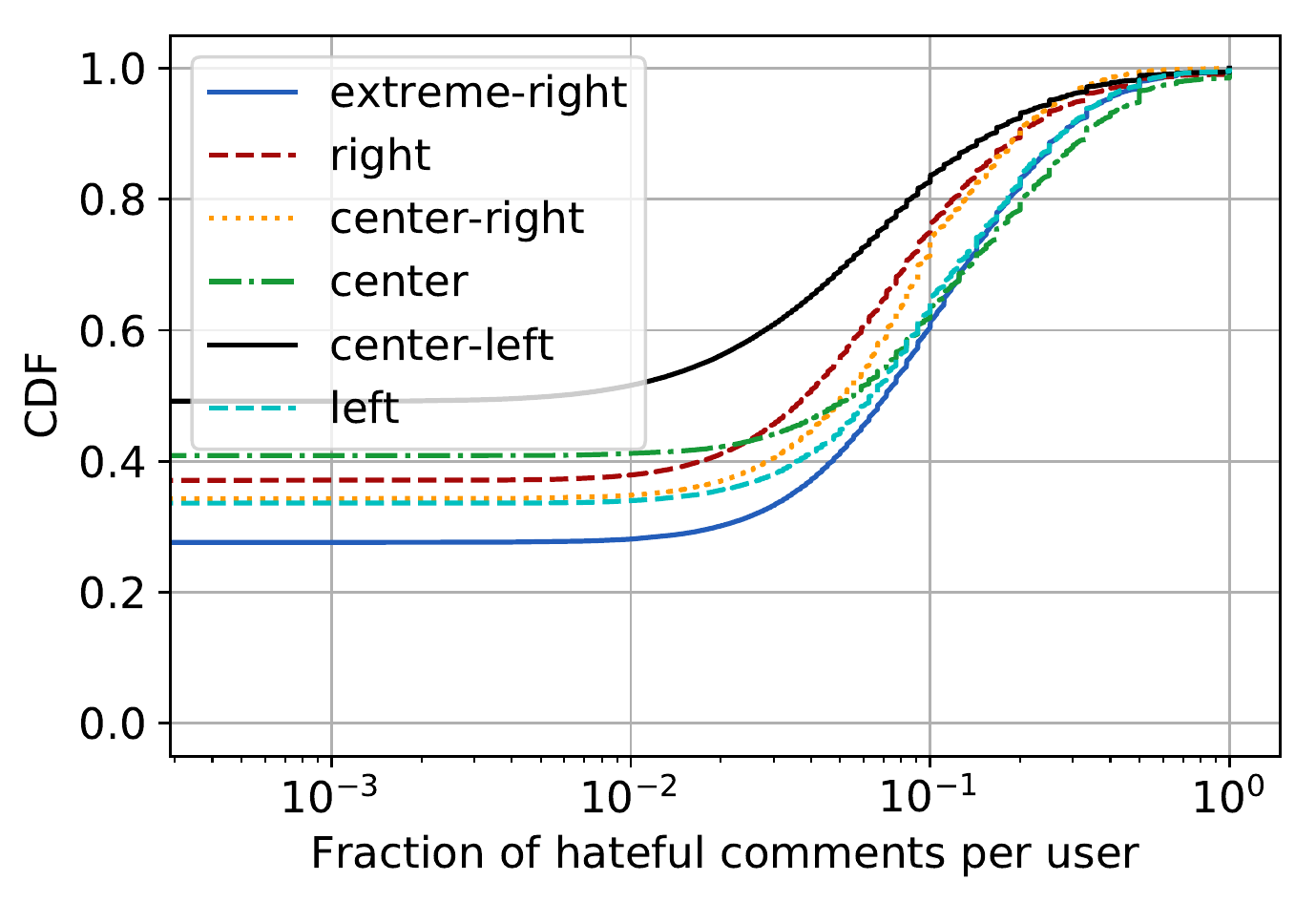}\label{subfig:cdf_fraction_hateful_comments_per_user_more10}}  
\caption{CDF of the fraction of hateful comments per user.}%
\label{fig:cdf_fraction_hateful_comments_per_user}
\end{figure}

Fig.~\ref{fig:cdf_fraction_hateful_comments_per_user} shows the fraction of hateful comments over all the comments that a user made per partisanship.
We make several observations.
First, a large percentage of users across all partisanships post only non-hateful comments: e.g., for extreme-right 56\% of the users post only non-hateful comments, while for other partisanships like center-right and center-left the percentage is much higher reaching 84\%.
When we look at the results for the users with at least ten comments (see Fig.~\ref{subfig:cdf_fraction_hateful_comments_per_user_more10}), however, we note that these percentages are substantially lower compared to all users. 
This indicates that ``power-users'' are more likely to share hateful comments, while users that are posting only a few times are less likely to post hateful comments.
Second, we note that users that post on extreme-right and right news articles are more likely to post hateful comments compared to users active on center- or left- leaning news articles.

\begin{figure}[t!]
\center
\subfigure[All users]{\includegraphics[width=0.8\columnwidth]{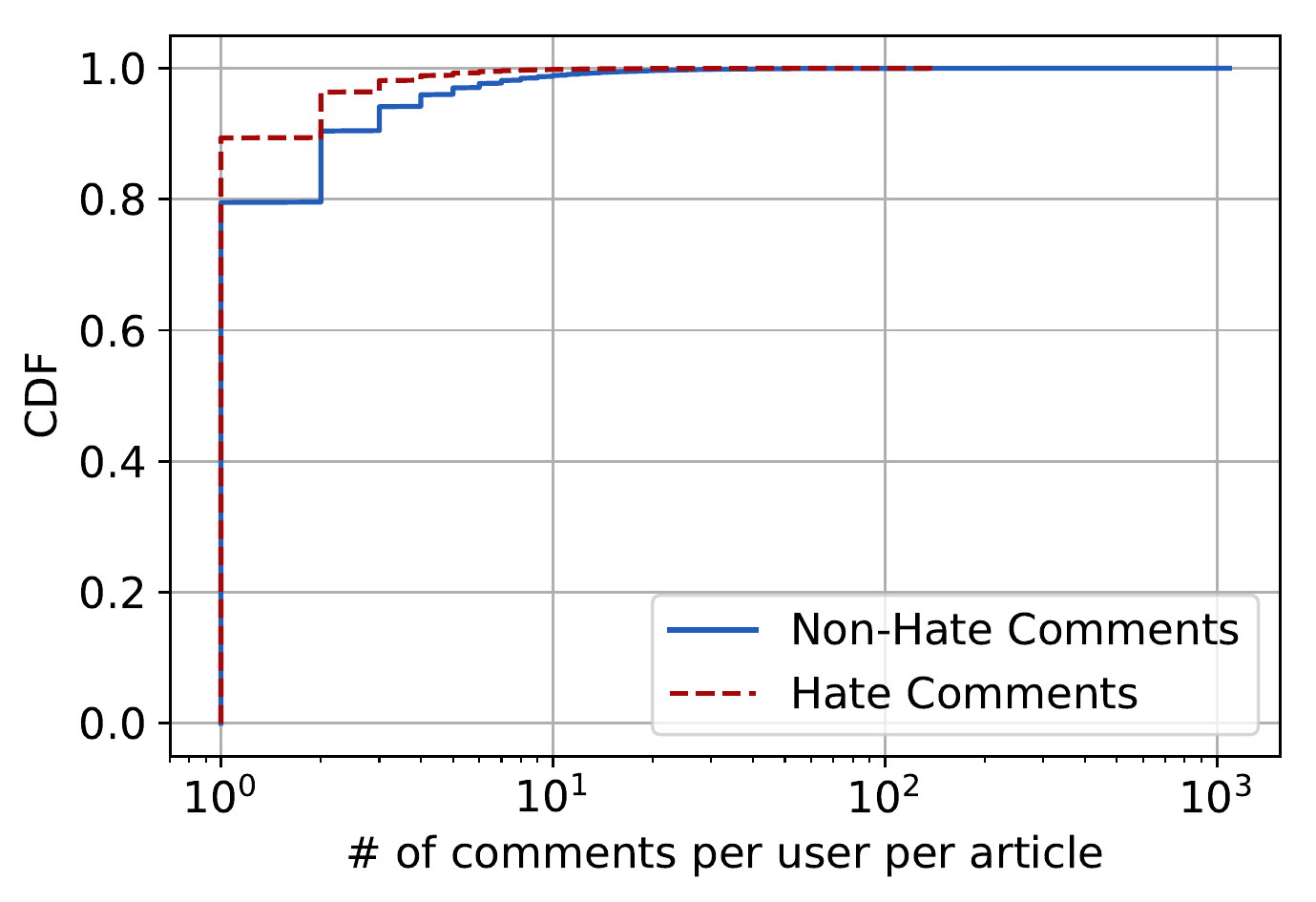}\label{subfig:cdf_comments_per_user_per_article}}
\subfigure[Users~with~at~least~ten~comments]{\includegraphics[width=0.8\columnwidth]{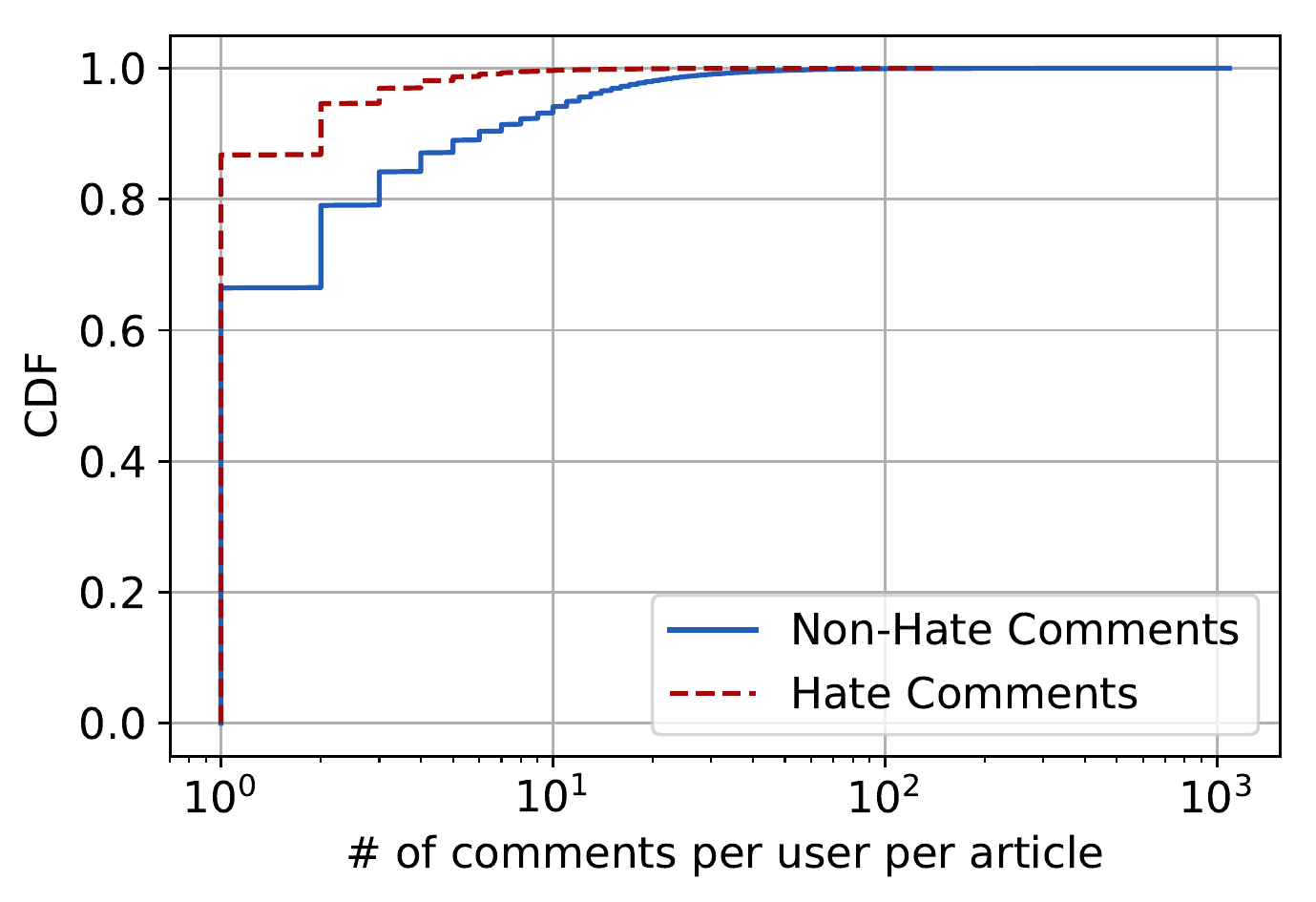}\label{subfig:cdf_comments_per_user_per_article_more10}}  
\caption{CDF of the number of comments per user per article.}%
\label{fig:cdf_comments_per_user_per_article}
\end{figure}

\descr{User Activity per Article.} Finally, we analyze the user commenting activity in the granularity of specific articles. This analysis allows us to understand the discussion on specific news articles and whether users that post hateful comments are persistent (i.e., posting multiple hateful comments) or whether they are ``one-off.'' 
We plot the CDF of the number of comments per user for each article by distinguishing between hateful and non-hateful comments in Fig.~\ref{fig:cdf_comments_per_user_per_article}.
We observe that for both hateful and non-hateful comments, a large percentage of users post only once on the news article.
This happens for 79\% for non-hateful comments and 89\% for hateful comments, while by only considering users that posted over ten times (see Fig.~\ref{subfig:cdf_comments_per_user_per_article_more10}) the percentages decline to 66\% for non-hateful and 86\% for hateful comments.
Also, we run a KS test on the distributions in Fig.~\ref{fig:cdf_comments_per_user_per_article}, finding that the distributions exhibits statistically significant differences ($p<0.01$).
These results indicate that it is more likely that users that post non-hateful comments to hold a lengthy discussion on news articles, while users that post hateful comments are more likely to just post a single hateful comment once and then do not post other hateful comments. 
Note that we performed the same analysis by dividing the users according to their activity in news articles per partisanship finding no substantial differences between the results across partisanships (we omit these results from the manuscript).

\begin{figure*}[t!]
\center
\subfigure[Summary Scores]{\includegraphics[width=0.49\linewidth]{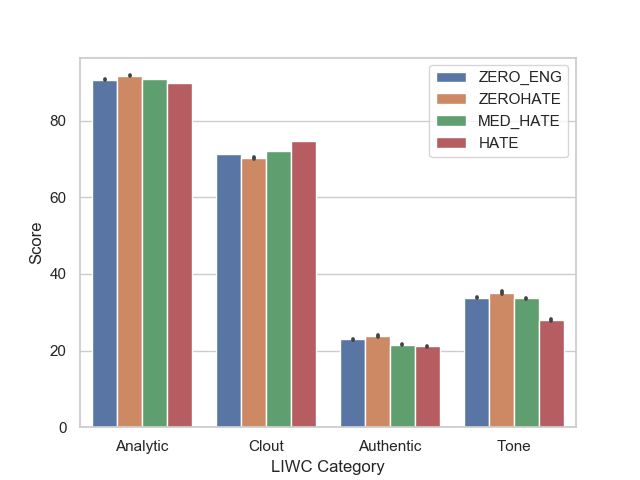}\label{subfig:hate_analytic_clout_authentic}}
\subfigure[Psychological Process]{\includegraphics[width=0.49\linewidth]{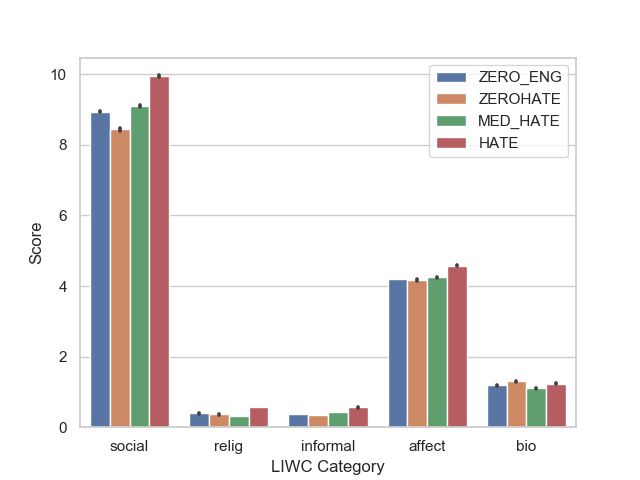}\label{subfig:hate_psychprocess}}
 \subfigure[Person pronouns]{\includegraphics[width=0.49\linewidth]{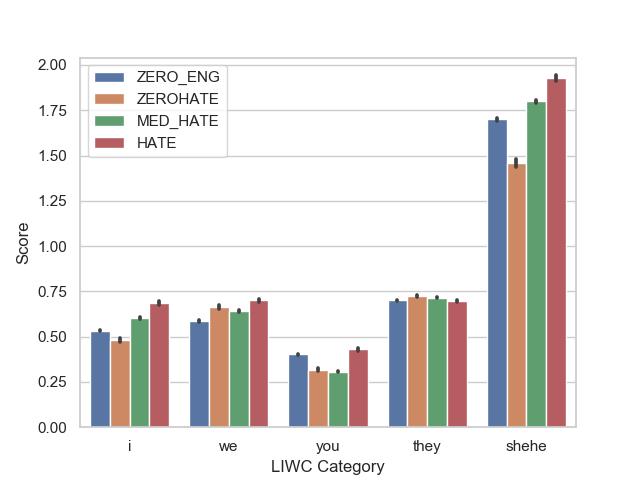}\label{subfig:hate_person_pronouns}}
 \subfigure[Negative Emotions]{\includegraphics[width=0.49\linewidth]{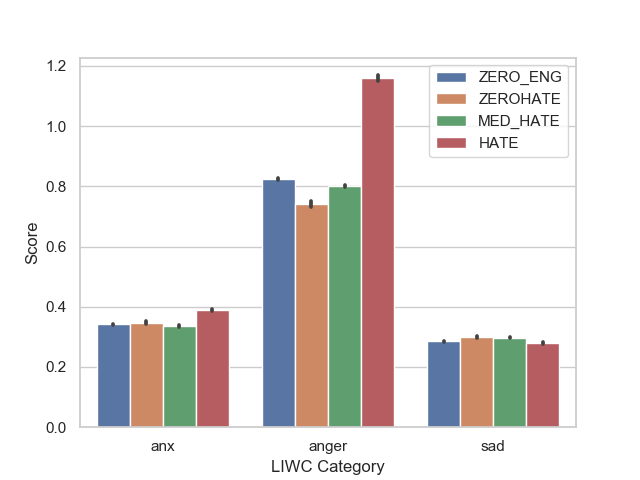}\label{subfig:hate_neg_emot}} 
  \subfigure[Cognitive processes]{\includegraphics[width=0.49\linewidth]{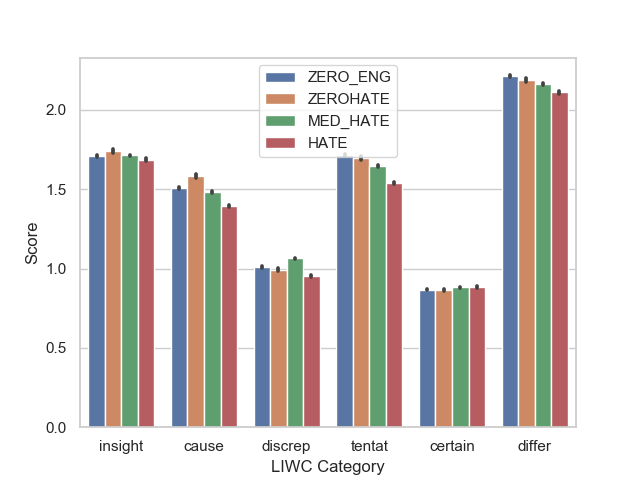}\label{subfig:hate_cog_proc}}
   \subfigure[Drives]{\includegraphics[width=0.49\linewidth]{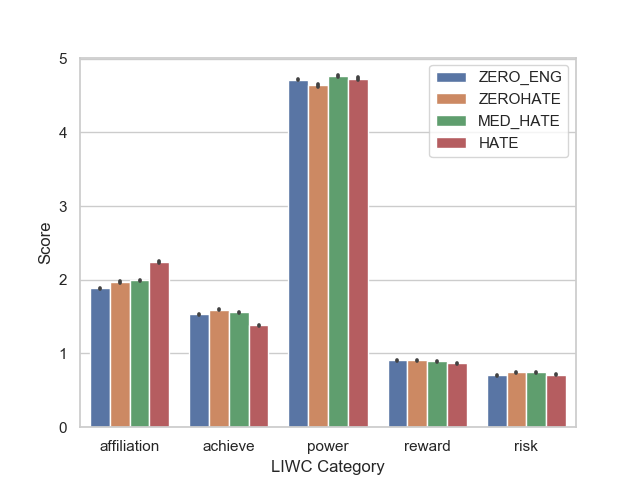}\label{subfig:hate_drives}} 
  \caption{Mean scores for LIWC categories across articles with different level of hate comment.}
\label{fig:liwc_results_hate}
\end{figure*}

\subsection{Content Linguistic Analysis}
In Journalism, extensive research have studied news article construction for better reader engagement~\cite{catenaccio2011towards, ha2018decline, o2011exploring, kim2013stumbling, yaros2006medium, mersey2012focusing, bell1991language}.
In this section, we assess whether specific linguistics used in news articles have any correlation with hate intensity. 
This analysis is important as it sheds light into the linguistics that drive hateful activity in news article comments. 
These cues can later be used to %
predict whether an article is likely to attract hate based on linguistics.

In our analysis, we divide the collection of news articles into four types of articles based on their comment engagement and hate intensity in their associated comments: first, those that do not receive any engagement in terms of number of comments (ZERO\_ENG); second, those that receive no hate comments (ZERO\_HATE); third, those for which the number of hateful comments exceeds a pre-defined threshold $k$ (HATE); and finally, the rest of the articles, which are the ones that receive at least one hate comment but less than the pre-defined threshold $k$ (MED\_HATE). 
By checking the CDF of the hate fraction in different articles (see Fig.~\ref{subfig:cdf_hate_comments_per_partisanship}), we observe that a threshold of 10\% over all comments represents a substantial number of articles; hence we set $k=10\%$. 
Using this threshold, we find that 52.4\% of the articles are ZERO\_ENG, 7.3\% are ZERO\_HATE, 33.2\% are MED\_HATE, 7.1\% are HATE articles. 
\descr{Articles' Linguistic Styles and Hate Comments.} 
The interplay of language use and journalism, media and society has been the focus of political science and journalism research~\cite{van2015between, lukin2013journalism}. 
In particular, many principles of journalism are grounded in psycho-linguistic
research, the study of how language is acquired, represented, and used~\cite{mcadams1984psycholinguistics}. %
To better understand the characteristics of the articles and their relation to receiving hate comments, we perform a psycho-linguistic analysis on the news articles. 
For a full psycho-linguistic analysis, we use a tool called Linguistic Inquiry and Word Count (LIWC)~\cite{chung-pennebaker}. LIWC is a text analysis program that calculates the degree to which various categories of words are used in a text. 
LIWC has been widely adopted by researchers to study emotional, cognitive, and structural components present in individuals' verbal and written speech samples.
We focus on the following dimensions provided by the tool: \emph{summary scores}, \emph{psychological processes}, and \emph{linguistic dimensions}.
\emph{Summary scores} include general attributes derived from the text, like the authenticity of the text, and basic statistics, like words per sentence. \emph{Psychological processes} describe the emotions that the text exposes, and \emph{linguistic attributes} describe the linguistic style of the text. %
We perform the analysis on each article. Fig.~\ref{fig:liwc_results_hate} shows the mean scores for our key LIWC attributes.
To assess the statistical significance of our results, we perform unpaired (two sample) t-tests with a 95\% confidence interval for the difference between the means.
Our analysis yields the following observations:

\noindent \textbf{HATE articles include content with the highest \emph{clout} scores and the least \emph{tone} scores in comparison to all other articles.}
Fig.~\ref{subfig:hate_analytic_clout_authentic} shows the language values obtained from LIWC averaged over all content for ZERO\_ENG, ZERO\_HATE, MED\_HATE, and HATE articles. 
We show that HATE articles have the highest mean ($\mu$ = 74.67, $p < 0.05$) for clout (influence and power) values and the lowest mean ($\mu$ = 28.06, $p < 0.05$) for tone. The high \emph{clout} score suggests that the linguistic style of HATE articles is associated with high expertise and confident cues, which can be used to influence an audience. 
Also, the low \emph{tone} scores suggest that the linguistic style of HATE articles is associated with the highest \emph{negative} tone. 
\noindent \textbf{HATE articles include content with the highest \emph{social}, \emph{religion,} and \emph{affect} references in comparison to all other articles.}
 Fig.~\ref{subfig:hate_psychprocess} shows that HATE articles have the highest mean for the social ($\mu$ = 9.94, $p < 0.05$), religion ($\mu$ = 0.57 , $p < 0.05$), and affect ($\mu$ = 0.56 , $p < 0.05$). 
 \emph{Social processes} include family, friends, female and male references. For example, an excerpt from a news article, belonging to the HATE category, that evokes the social category is ``\textit{Hillary Clinton has an explanation for why women white women in particular voted against her last November they caved in to pressure from their husbands fathers boyfriends and male bosses}.'' Our analysis also reveals that HATE articles reference religion-related entities and are on average more emotional than other types of articles.

\noindent \textbf{On average, HATE articles include the highest first (\emph{I}) and third person (\emph{she/he}) singular pronouns in comparison to all other types of articles.}
 Fig.~\ref{subfig:hate_person_pronouns} shows that HATE articles have the highest mean for scores associated with first ($\mu$ = 0.68, $p < 0.05$) and third singular pronouns ($\mu$ = 1.92 $p < 0.05$).
These findings show that articles which are about individual people, or include and cite their opinions receive hate comments with higher probability.
\noindent \textbf{HATE articles include the highest \emph{anger} and \emph{anxiety} references.} 
 Fig.~\ref{subfig:hate_neg_emot} shows that \emph{anger} is the most prevalent negative emotion for all three types of articles. In particular, HATE articles on average have the highest level of anger ($\mu$ = 1.15, $p < 0.05$). 
 Also, we find that HATE articles on average have the highest level of anxiety ($\mu = 0.39$, $p < 0.05$).

 Emotion and journalism have already been well studied~\cite{wahl2016emotion, beckett2016role,pantti2010value, newhagen1992evening, palazzolo2011media}. Mostly, the focus is on how to use emotion to have quality reporting and editing, and to articulate the news more effectively. The use of emotion for manipulation has also been studied~\cite{valentino2008worried, richards2007emotional, valentino2011election}. Moreover, findings in political psychology suggest that specific emotions may play an important role in political mobilization. Our finding in particular is aligned with others who also identified anger, more than anxiety or enthusiasm, to mobilize~\cite{valentino2011election}. 
\noindent \textbf{HATE articles include the least number of words that suggest causation, discrepancy, tentative, and differentiation.}
Fig.~\ref{subfig:hate_cog_proc} shows that HATE articles tend to have the lowest scores for causation ($\mu$ = 1.39,  $p < 0.05$), discrepancy (words like ``would'' and ``should,'' $\mu$ = 0.95,  $p < 0.05$), tentative (words like ``maybe'' and ``perhaps,'' $\mu$ = 1.53,  $p < 0.05$), and differentiation (words like ``hasn't,'' ``but,'' and ``else,'' $\mu$ = 2.1,  $p < 0.05$).  This can indicate that HATE articles tend to have less justification of arguments in terms of causes or effects.

\noindent \textbf{HATE articles include the highest references related to affiliation and the lowest references to achievement.} Fig.~\ref{subfig:hate_drives} shows that HATE articles have the highest mean for words suggesting affiliation ($\mu$ = 2.23, $p < 0.05$) and the lowest achievement references ($\mu$ = 1.38 , $p < 0.05$). 
This likely suggests that HATE articles are motivated by the need to be affiliated to certain groups and because of their negative nature they might not mention achievements.

We also perform the linguistics analysis on the articles grouped by partisanship to understand if news sites with different partisanship have differences in terms of linguistic dimensions.
In general, we find minor differences in the linguistic dimensions across partisanship with some exceptions:
articles that have a center partisanship have the least affect and the least focus on the past in comparison with articles from other partisanship.

\begin{figure}[t!]
\center
\subfigure[All comments]{\includegraphics[width=\columnwidth]{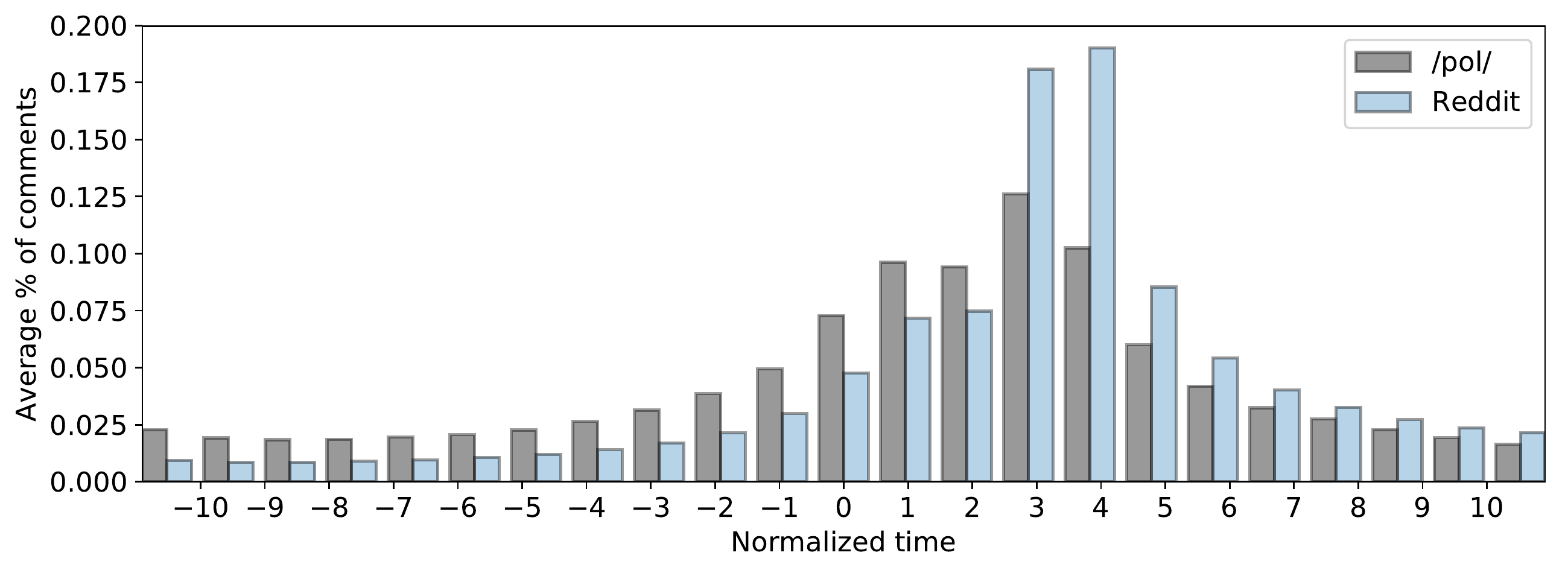}\label{subfig:temporal_hist_all_comments}}
\subfigure[Hate comments]{\includegraphics[width=\columnwidth]{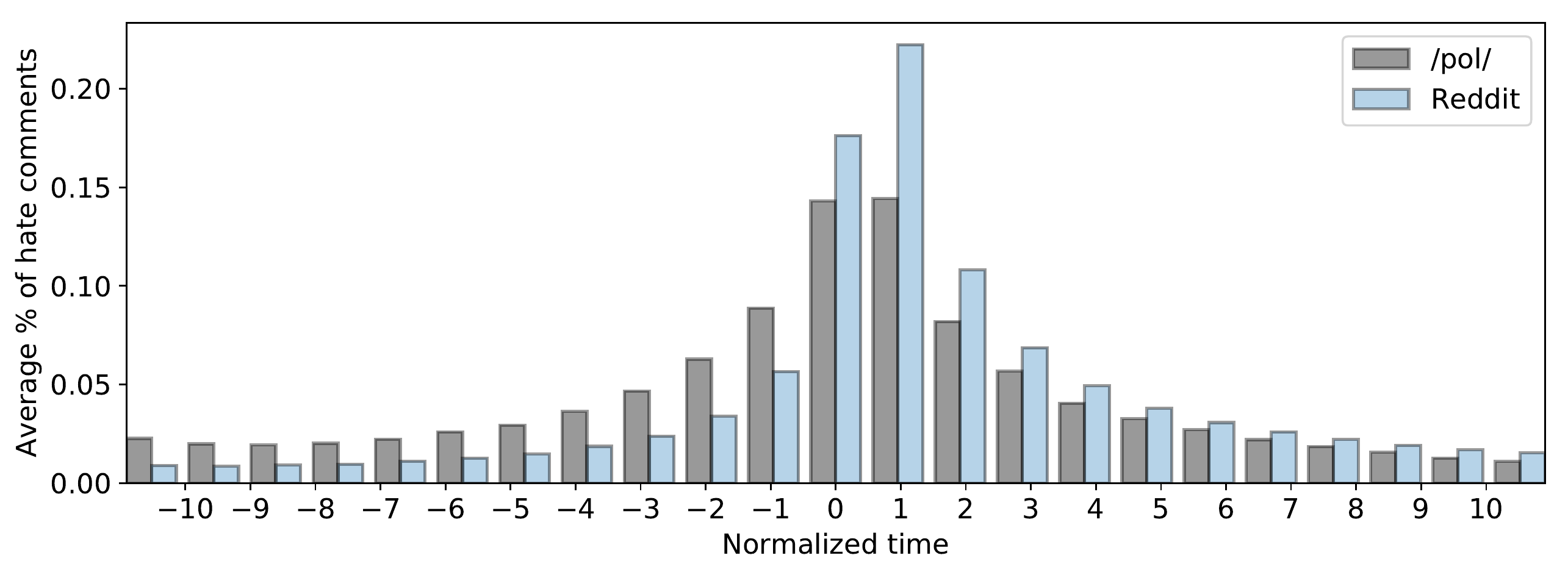}\label{subfig:temporal_hist_hate_comments}}
\subfigure[Hate comments/All comments]{\includegraphics[width=\columnwidth]{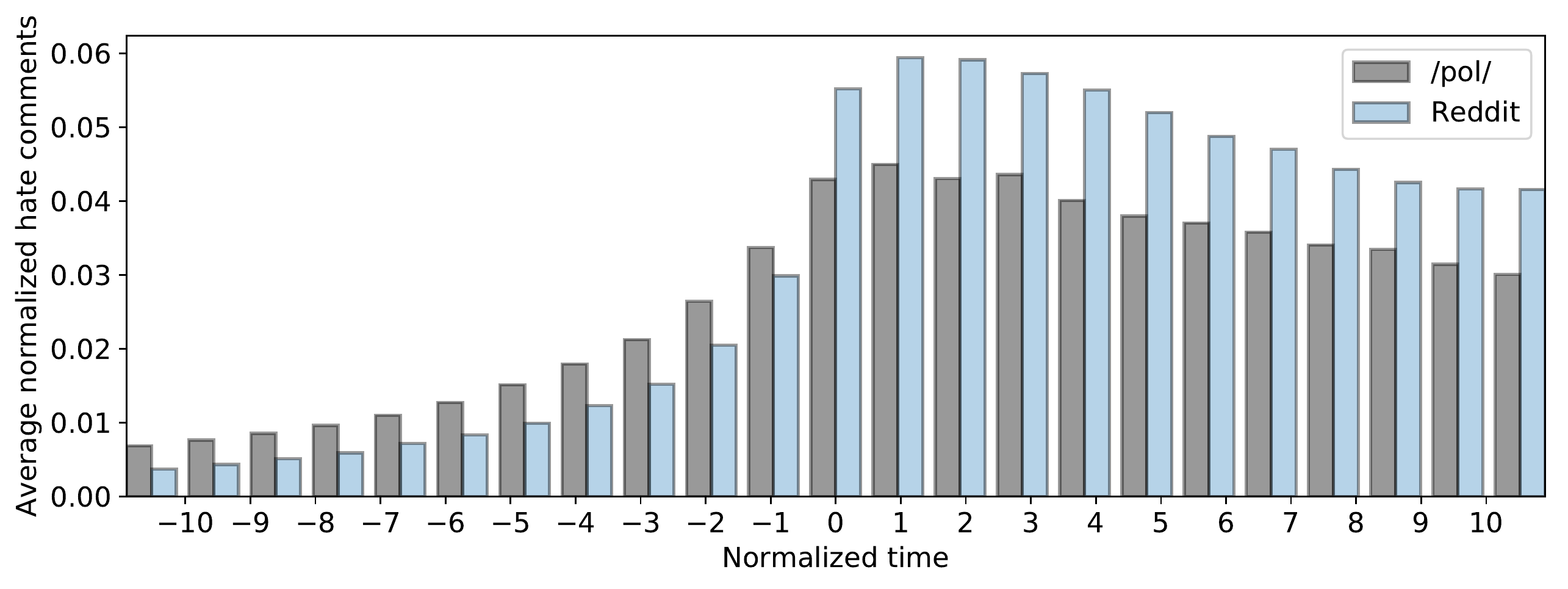}\label{subfig:temporal_hist_norm_comments}}
\caption{Increase of comment activity over time after the post of news articles on six subreddits or /pol/.}
\label{fig:temporal_hist_all_domains}
\end{figure}

\begin{figure*}[t!]
\center
\subfigure[Left]{\includegraphics[width=0.49\textwidth]{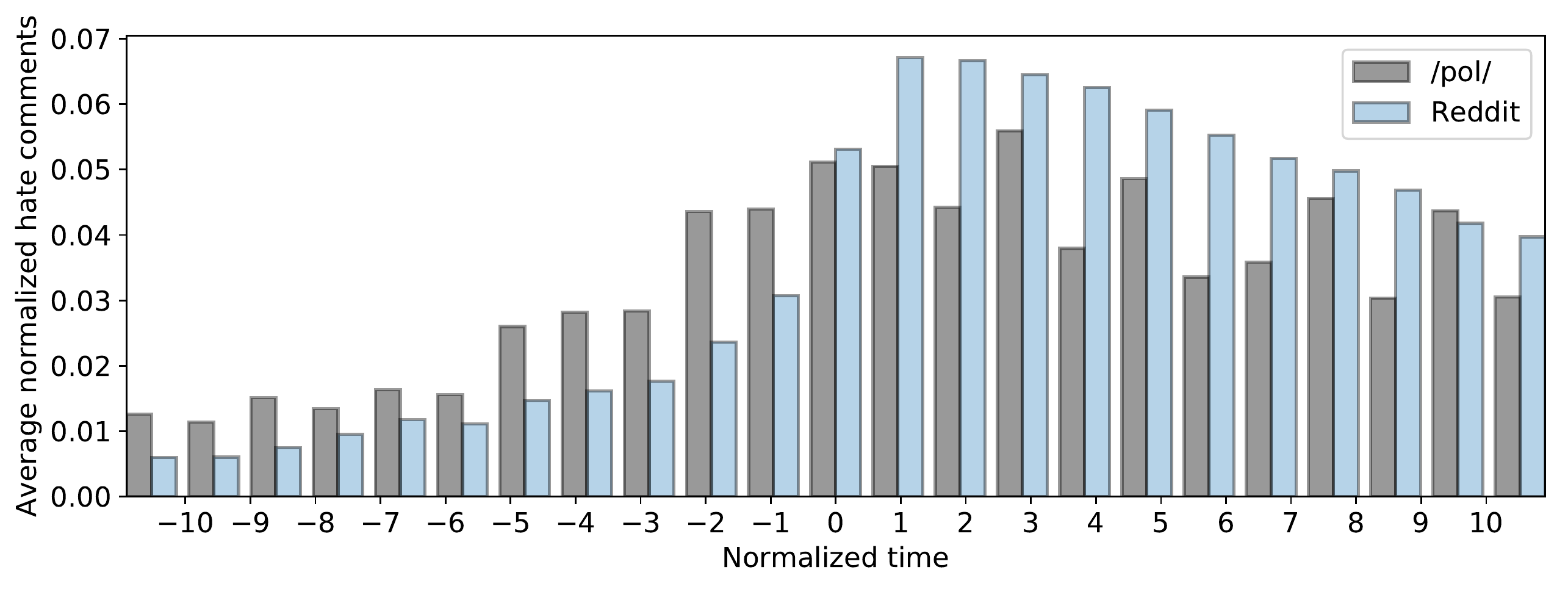}\label{subfig:temporal_hist_hate_part_left_norm}}
\subfigure[Center-Left]{\includegraphics[width=0.49\textwidth]{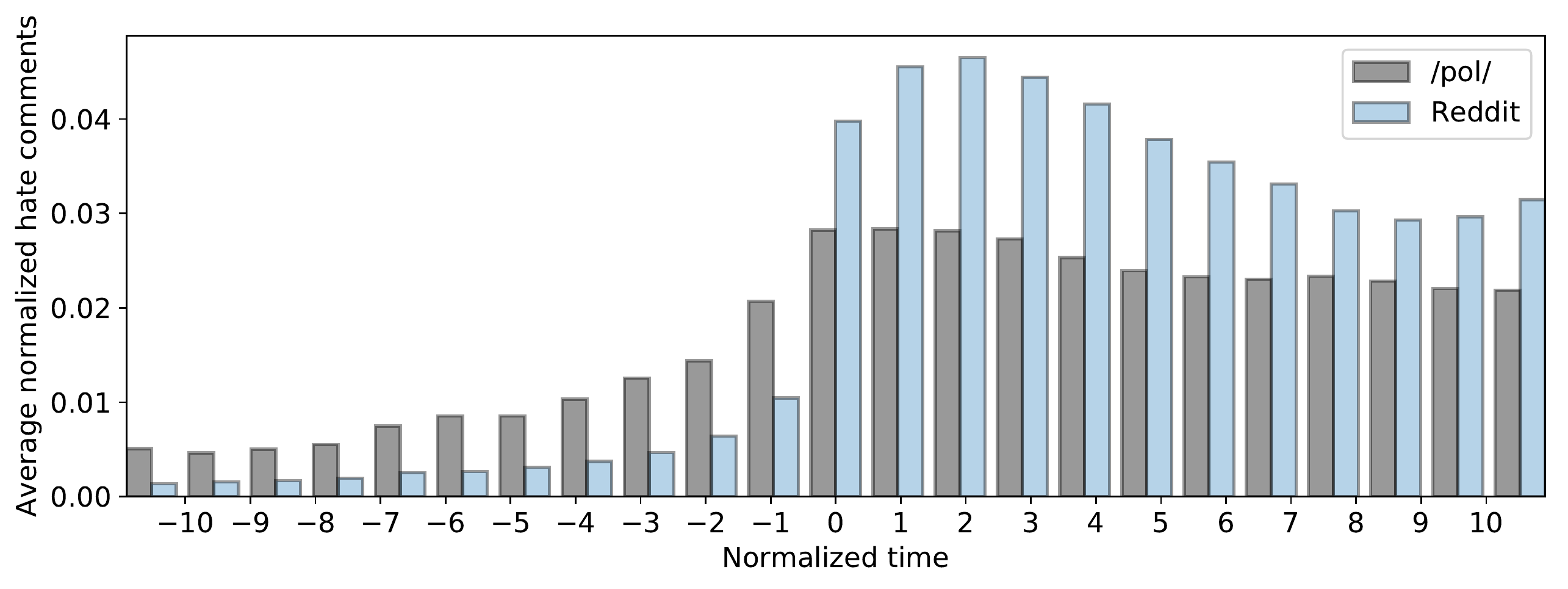}\label{subfig:temporal_hist_hate_part_center_left_norm}}
\subfigure[Center]{\includegraphics[width=0.49\textwidth]{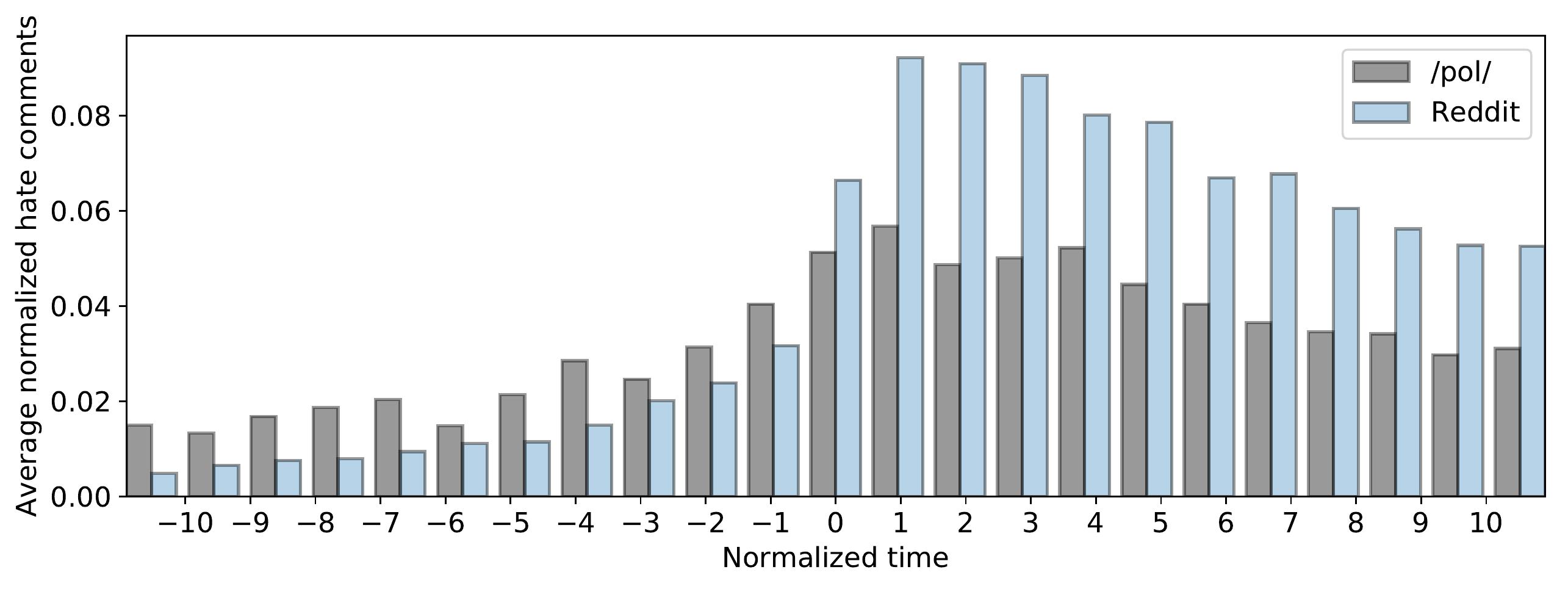}\label{subfig:temporal_hist_hate_part_center_norm}}
\subfigure[Center-Right]{\includegraphics[width=0.49\textwidth]{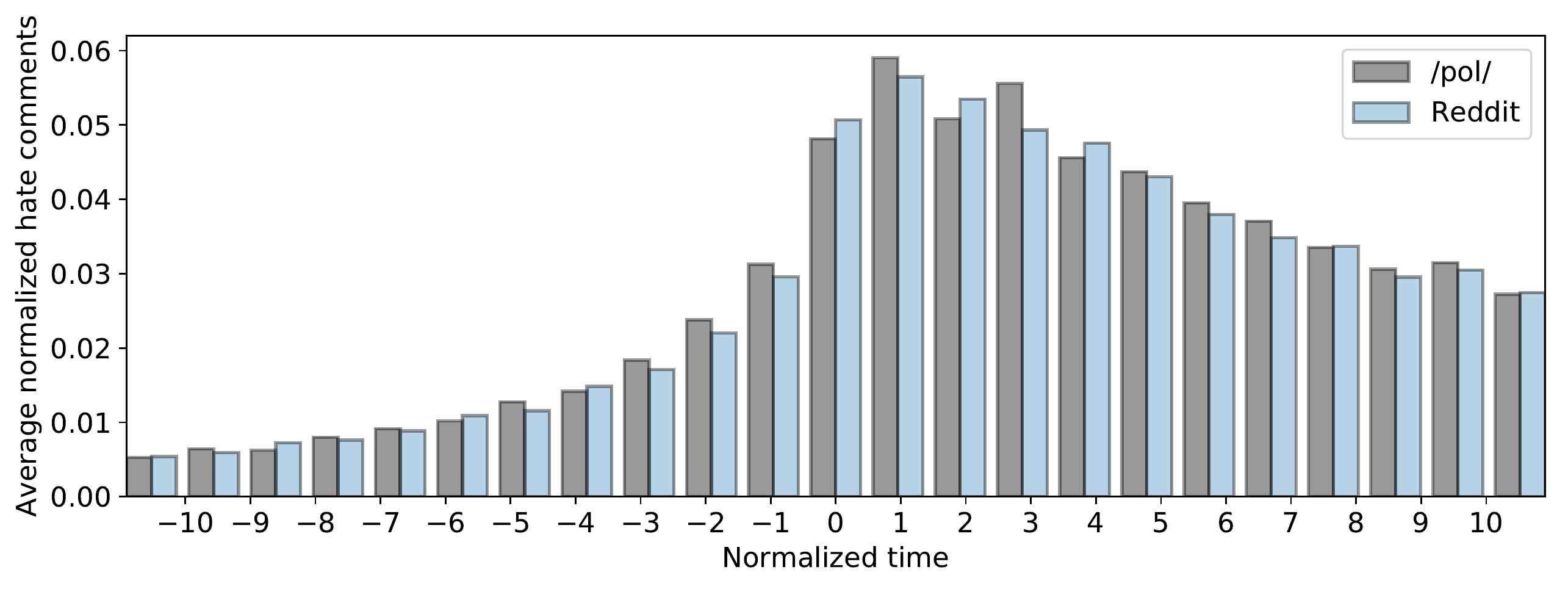}\label{subfig:temporal_hist_hate_part_center_right_norm}}
\subfigure[Right]{\includegraphics[width=0.49\textwidth]{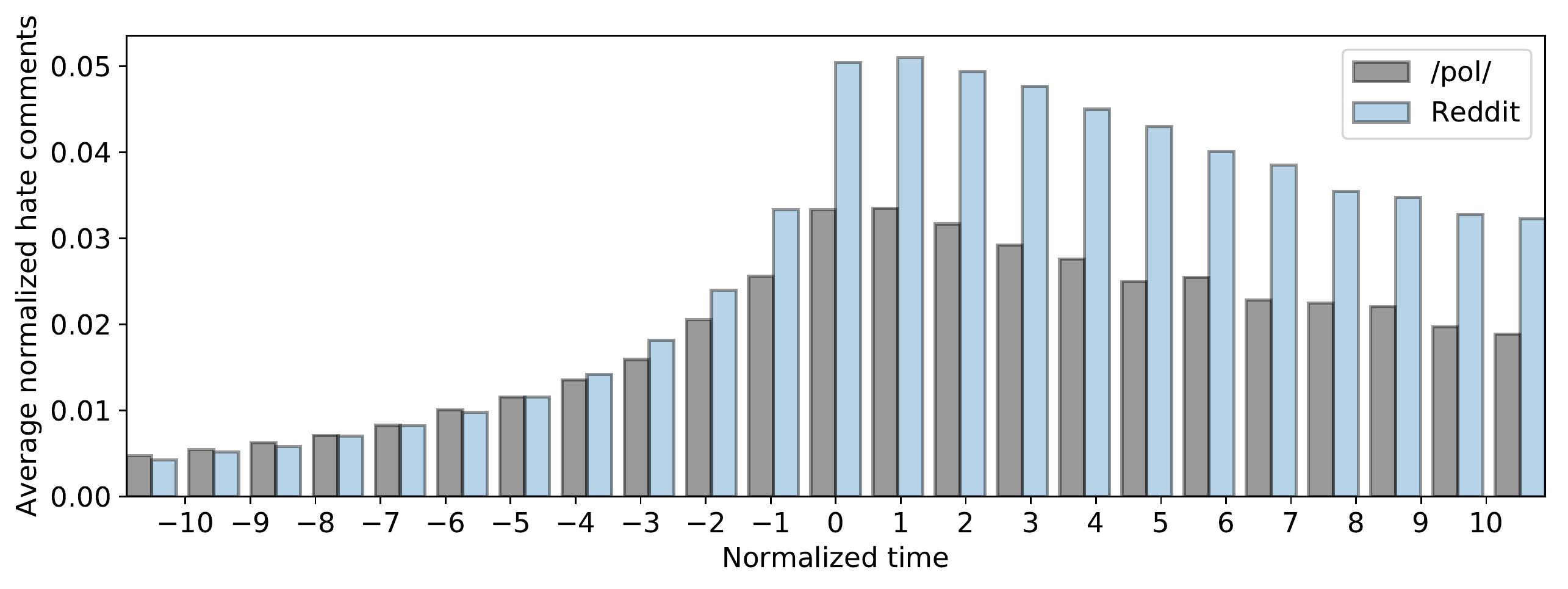}\label{subfig:temporal_hist_hate_part_right_norm}}
\subfigure[Extreme Right]{\includegraphics[width=0.49\textwidth]{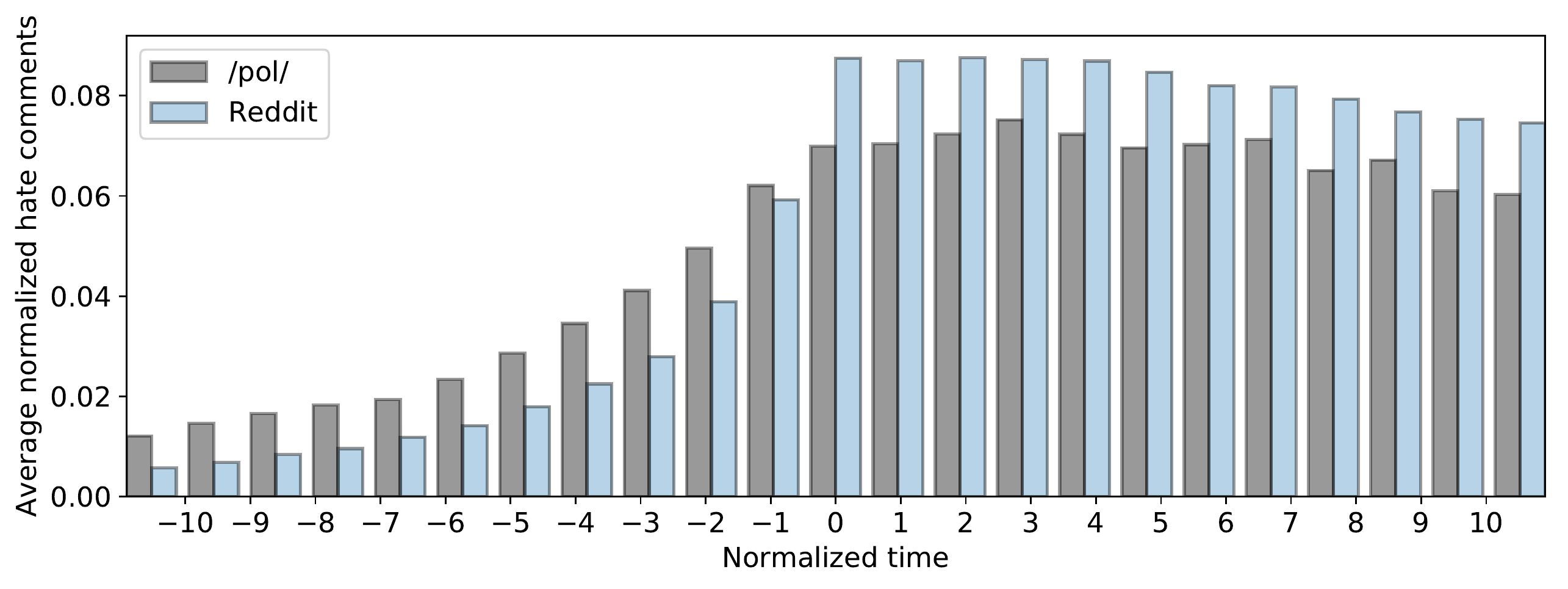}\label{subfig:temporal_hist_hate_part_extreme_right_norm}}
\caption{Fraction of hate comments over all comments for each normalized timeslot.} 
\label{fig:temporal_hist_partisanship_norm}
\end{figure*}

\begin{figure}[t!]
\center
\subfigure[All comments]{\includegraphics[width=\columnwidth]{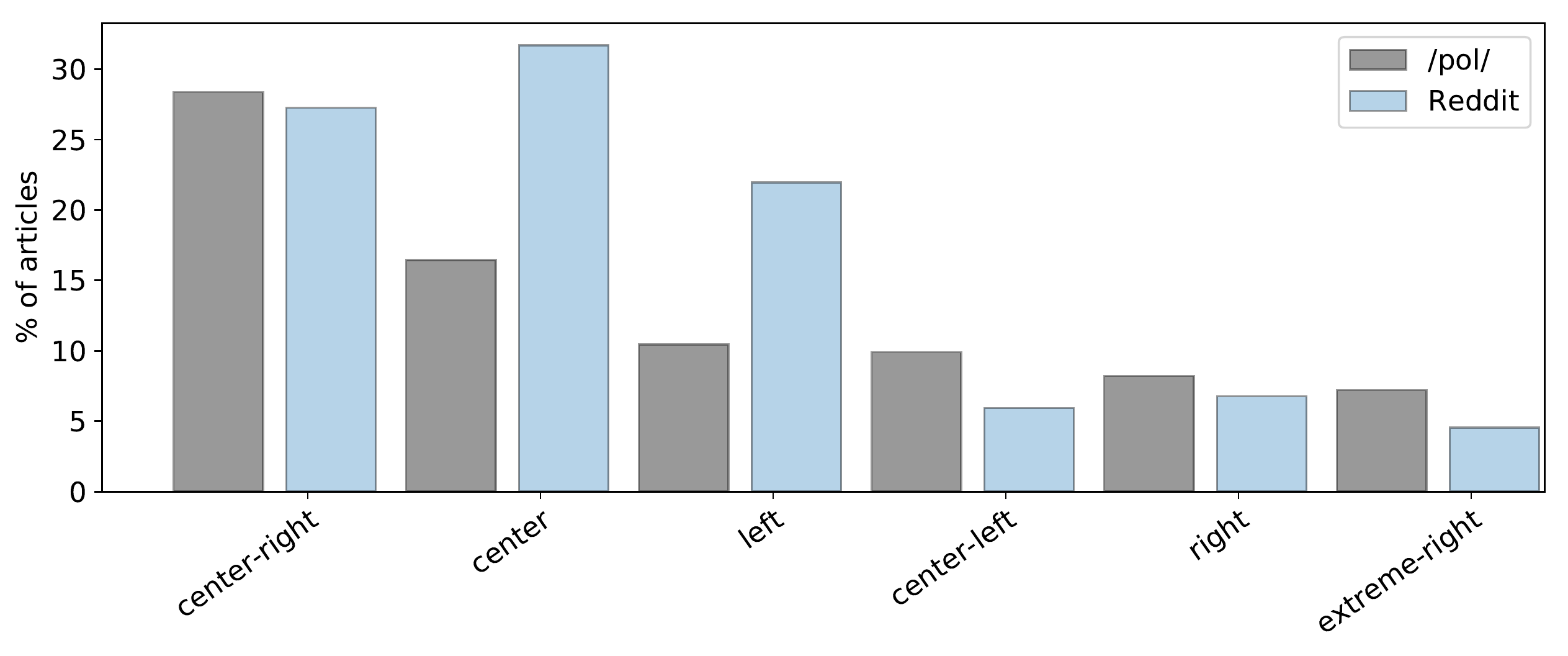}\label{subfig:bc_percentage_interesting_partisanship}}
\subfigure[Hate comments]{\includegraphics[width=\columnwidth]{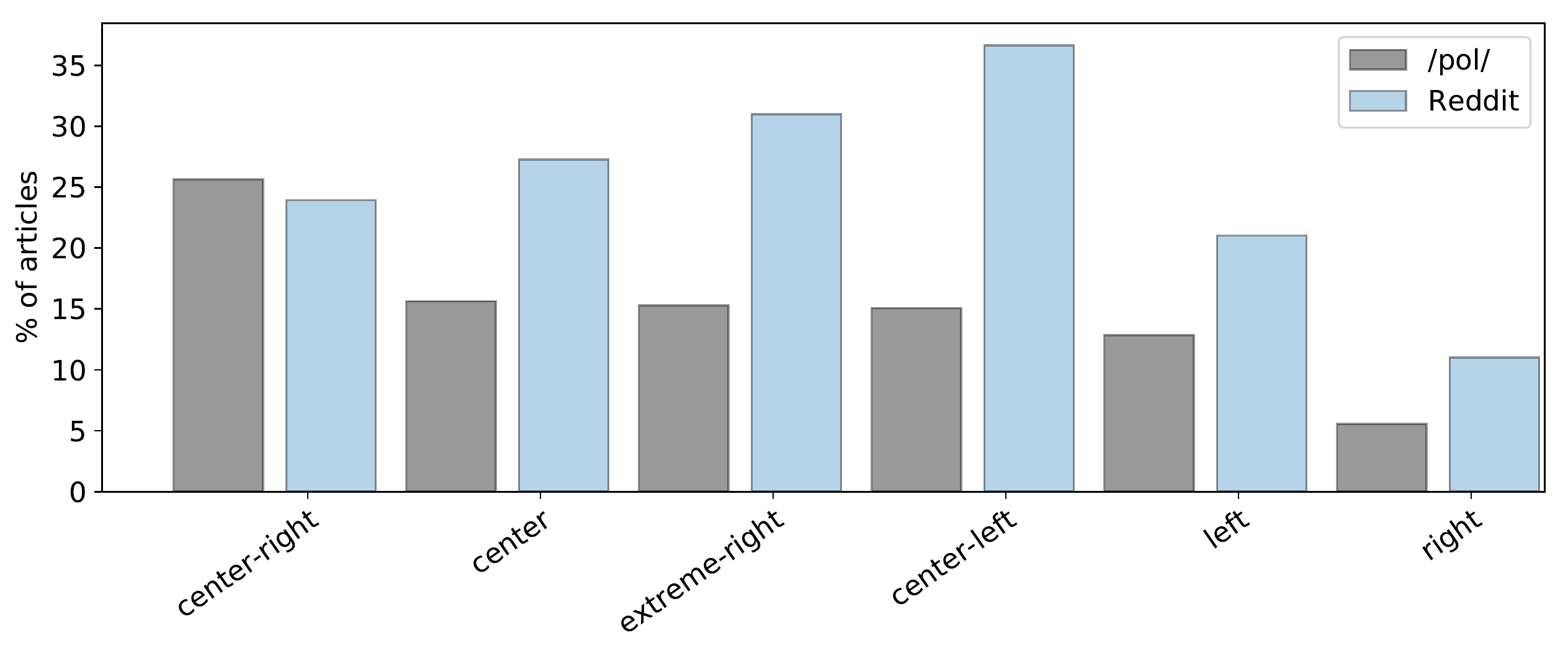}\label{subfig:bc_percentage_interesting__partisanship_hate}}
\caption{Percentage of news articles with increased commenting activity after appearing on /pol/ and/or the six subreddits.}
\label{fig:bc_percentage_interesting}
\end{figure}

\subsection{Activity after Social Network Posts}

In this section, we study the commenting activity on news articles after they appear on social networks.
We aim to provide answers to the following questions:
1) Is the appearance of news articles on social networks like 4chan and Reddit correlated with the (hateful) commenting activity on news articles? 
2) How does the (hateful) commenting activity decay after the posting of news articles on 4chan and Reddit? 
3) What portion of news articles receive increased hateful activity shortly after appearing in other social networks?
This analysis is important since it sheds light into the external factors (\emph{i.e.,} appearance of news articles on other social networks) that possibly affect the commenting activity on news sites.

To provide answers to the above questions, we find the first occurrence of each news article on the six subreddits and on /pol/.
Then, we normalize the occurrence of each comment in the news article, with respect to the first occurrence of the article in each platform, hence obtaining a view of whether comments, and in particular hate comments, increase after the appearance of articles on Reddit and 4chan.
To do this, we subtract the timestamp of each comment in news articles with the timestamp of the first occurrence of the article on the six subreddits and /pol/, hence obtaining a normalized time for the comments. 
Fig.~\ref{fig:temporal_hist_all_domains} shows the average percentage of comments that were posted in close proximity with the first occurrence of each article on the six subreddits and /pol/. 
Time zero corresponds to the first occurrence of the article on /pol/ or the six subreddits, while each bar corresponds to a time period of two hours.
For instance, the bars that have the number zero correspond to the time interval between the first occurrence of the article and the next two hours.
We report the results using three ways: Fig.~\ref{subfig:temporal_hist_all_comments} shows the occurrence of all comments per normalized time slot, Fig.~\ref{subfig:temporal_hist_hate_comments} shows the occurrence of hateful comments per normalized time slot, while Fig.~\ref{subfig:temporal_hist_norm_comments} shows the fraction of hateful comments over all comments per normalized time slot. The latter is useful as it captures the correlation between the hateful commenting activity and the overall activity.

We observe that for all comments (see Fig.~\ref{subfig:temporal_hist_all_comments}) the commenting activity increases after the first occurrence of the news articles in the six subreddits and /pol/ (normalized time 0) with a peak of activity at normalized time 3 and 4 for /pol/ and the six subreddits, respectively.
Also, we find that the commenting activity close to the first occurrence (between 0 and 2 normalized time) is greater for /pol/ when compared to the six subreddits, while later on (after normalized time 2) the percentage activity is larger for the six subreddits.
This is likely due to Reddit bots that post news articles without user interaction and likely because of 4chan's ephemeral nature: 4chan users are more likely to interact with the article closer to the article's post on the platform, as threads are short-lived. 
By only considering the hateful commenting activity (see Fig.~\ref{subfig:temporal_hist_hate_comments}), we observe a similar pattern with the important difference that the peak in hateful activity is closer to the appearance of the articles on the six subreddits and /pol/, namely during normalized time 1. 
This indicates that hateful commenting activity increases substantially right after the appearance of news articles on the six subreddits and /pol/, in a far quicker pace when compared to the overall commenting activity.

To further study the interplay between the overall commenting activity and the hateful commenting activity, we plot the fraction of hate comments over all comments per normalized time in Fig.~\ref{subfig:temporal_hist_norm_comments}. 
We observe that despite the fact that the overall commenting activity and hateful activity decreases substantially after normalized time 4 (see Fig.~\ref{subfig:temporal_hist_all_comments} and Fig.~\ref{subfig:temporal_hist_hate_comments}) the fraction of hateful comments over all comments decreases more gradually and it remains close to the peak (normalized time 4) even at normalized time 10.
These results highlight that the hateful commenting activity remains high relative to the overall commenting activity in an article for a long period after the appearance of news articles on the six subreddits and/or 4chan's /pol/, hence indicating that once a news article receives substantial amount of hate it continues to receive a relatively high fraction of hateful comments for a long time period.

Next, we make the same analysis focusing on hate comments, by grouping the articles according to each news site's partisanship (see Table~\ref{tbl:partisanships}).
Fig.~\ref{fig:temporal_hist_partisanship_norm} shows the fraction of hateful comments over all comments per normalized time period for each partisanship (we omit the figures for the overall commenting activity and overall hateful commenting activity due to space constraints).
We find that extreme-right news sites are more persistent in hateful commenting activity as the fraction of hateful comments over all comments decays substantially slower compared to the other partisanships. 
On the other hand, news sites that are more on the center (\emph{i.e.,} center, center-left, center-right) have the fastest decay of hateful comments over all comments.
These findings indicate that extreme news sites (\emph{i.e.,} extreme-right) are more likely to maintain a substantial percentage of hateful commenting activity after the appearance of news articles on the six subreddits and /pol/ when compared to other partisanships on the center.

These results are based on all the articles in our dataset that have at least one comment. 
However, not all articles receive hate comments after their first occurrence in other platforms like /pol/ and the six subreddits.
To understand this phenomenon and its prevalence on the Web, we filter the articles so that we select the ones that had the maximum (hateful) commenting activity during the normalized time zero: we find 39K articles for hateful commenting activity and 17K for all commenting activity. 
Fig.~\ref{fig:bc_percentage_interesting} reports the percentage of articles over all articles (with at least one comment) that have an increase in commenting activity, and in particular hate commenting activity, shortly after the first occurrence of the news articles on /pol/ or the six subreddits. 
We find that domains that are \emph{center-right} have the most articles with commenting activity increase, while \emph{extreme-right} domains have the least (see Fig.~\ref{subfig:bc_percentage_interesting_partisanship}).
When considering only hateful activity (see Fig.~\ref{subfig:bc_percentage_interesting__partisanship_hate}), we find something similar: again, \emph{center-right} domains have the most articles with activity increase and in this case it is hateful. 
A possible explanation is that users from the six subreddits or /pol/ disagree or have a different ideology with articles from center-right news sites, hence posting hateful comments in the comments section right after their appearance on their platform.
Finally, we note that for hateful commenting activity the percentages are higher for Reddit across all partisanships with the exception of center-right, possibly indicating that Reddit users are more likely to post hateful comments on these news articles in close temporal proximity after their appearance on the six subreddits.

\section{Conclusion}
In this paper, we presented a large-scale quantitative analysis of the news commenting ecosystem.
We analyzed 125M comments and 412K news articles across several axes: we performed a general characterization of hateful content in news comments, a temporal analysis, as well as a linguistics characterization.
Overall, among other things, we found that (hateful) commenting activity increases with notable events that have a strong political nature, articles that attract varying hateful activity have significant linguistic differences, while our user-based analysis reveals that users that post comments in extreme-right sites tend to be more active and post more hateful comments compared to users that post on sites with other partisanships.
Furthermore, we found a correlation between the posting of news articles on either /pol/ or the six selected subreddits and increased (hateful) commenting activity on the article.

Naturally our work has some limitations. 
First, our dataset was collected well after the publication of the articles and their comments, hence it is likely that some of the hateful content was moderated/deleted.
Second, we relied on the Perspective API for detecting hate speech, which is expected to miss some hateful content (as mentioned in Section~\ref{sec:method}).
This is because hate speech detection is an open research problem and available classifiers are unable to detect all possible types of hateful content.

To conclude, for our future work, we plan to work on pro-actively detecting organized campaigns, mainly from users of fringe Web communities, that aim to ``raid'' news articles with hate comments.
Also, we aim to assess the effect that other mainstream social networks (e.g., Twitter) have on the commenting activity of news articles.
Finally, we plan to build a classifier that will be able to detect whether news articles are likely to attract hateful comments.

\descr{Acknowledgments.} This work was partially supported by the National Science Foundation (NSF) under Grant CNS-1942610.
Any opinions, findings, and conclusions expressed in this paper are those of the authors and do not necessarily reflect the views of the NSF.

\bibliographystyle{abbrv}

%

\end{document}